\documentclass[
	a4paper, % Paper size, use either a4paper or letterpaper
	10pt, % Default font size, can also use 11pt or 12pt, although this is not recommended
	unnumberedsections, % Comment to enable section numbering
	twoside, % Two side traditional mode where headers and footers change between odd and even pages, comment this option to make them fixed
]{LTJournalArticle}

\usepackage{bm}
\usepackage{comment}
\usepackage{biblatex}

\usepackage{amssymb}
\usepackage{latexsym}
\usepackage{url}
\usepackage{xcolor}

\usepackage{hyperref}

\usepackage{booktabs}       % professional-quality tables
\usepackage{amsfonts}       % blackboard math symbols
\usepackage{nicefrac}       % compact symbols for 1/2, etc.
\usepackage{microtype}      % microtypography
\usepackage{listings}
\usepackage{graphicx}
\usepackage{mathtools,eqparbox}
\usepackage{tabularx}
\usepackage{caption} 
\definecolor{newcolor}{rgb}{.8,.349,.1}

\title{ComBAT Harmonization for diffusion MRI: Challenges and Best Practices}%
\author{%
	Pierre-Marc Jodoin\textsuperscript{1,2}\thanks{Corresponding author: \href{mailto:pierre-marc.jodoin@usherbrooke.ca}{jane@smith.com}}, 
    Manon Edde\textsuperscript{1,3}  
    Gabriel Girard\textsuperscript{1}  
    Félix Dumais\textsuperscript{1}  \\
    Guillaume Theaud\textsuperscript{2}  
    Matthieu Dumont\textsuperscript{2}  
    Jean-Christophe Houde\textsuperscript{2}\\ 
    Yoan David\textsuperscript{1} 
    Maxime Descoteaux\textsuperscript{1,2} 
    Alzheimer’s Disease Neuroimaging Initiative\textsuperscript{4}
}
\date{\footnotesize\textsuperscript{\textbf{1}}Université de Sherbrooke, 2500 Boul. de l'Université, J1K 2R1 Sherbrooke, Qc, Canada\\ \textsuperscript{\textbf{2}}Imeka Solutions inc, 195 Belvédère nord, J1H 4A7, Sherbrooke,  Qc, Canada\\ \textsuperscript{\textbf{3}}StoP-AD Centre, Douglas Mental Health University Institute, McGill University, Montréal,H4H 1R3,  Quebec, Canada\\
\textsuperscript{\textbf{4}}  Data used in preparation of this article were obtained from the ADNI database (adni.loni.usc.edu). }

\renewcommand{\maketitlehookd}{%
	\begin{abstract}
		\noindent Over the years, ComBAT has become the standard method for harmonizing MRI-derived measurements, with its ability to compensate for site-related additive and multiplicative biases while preserving biological variability. However, ComBAT relies on a set of assumptions that, when violated, can result in flawed harmonization. In this paper, we thoroughly review ComBAT's mathematical foundation, outlining these assumptions, and exploring their implications for the demographic composition necessary for optimal results. 

Through a series of experiments involving a slightly modified version of ComBAT called {\em Pairwise-ComBAT} tailored for normative modeling applications, we assess the impact of various population characteristics, including population size, age distribution, the absence of certain covariates, and the magnitude of additive and multiplicative factors. Based on these experiments, we present five essential recommendations that should be carefully considered to enhance consistency and supporting reproducibility, two essential factors for open science, collaborative research, and real-life clinical deployment.\\
	\end{abstract}
}

\begin{document}

%\verso{Pierre-Marc Jodoin \textit{et~al.}}

%\begin{frontmatter}
\maketitle

\section{Introduction}
\label{sec:introduction}
In neuroimaging, leveraging multi-site data to analyze general trends and variability within diverse populations is increasingly common \cite{Di-Biase2023, Marquand2016, Rutherford2023, Verdi2021, Villalon-Reina2022}. However, variations in imaging protocols across sites introduce systematic biases, posing significant challenges to multi-site and longitudinal studies \cite{Cetin-Karayumak2024, Hu2023a, Huynh2019, Moyer2020, Schilling2021, Bayer2022,
Helmer2016, Pinto2020}. To address these biases, various harmonization techniques have been developed~\cite{Pinto2020, Chen2022, De-Luca2022, Horng2022, Hu2023a, Hu2023b, Johnson2007, Fortin2017}.  
These methods either pre-harmonize diffusion-weighted images via neural networks or other techniques like LinearRISH \cite{Moyer2020, Tax2019, Cetin-Karayumak2019} or post-harmonize MRI-derived measurements, such as through ComBAT~\cite{Johnson2007, Fortin2017, Fortin2018}.

ComBAT is one of the most effective methods (8,000+ citations as of March 2025 according to PubMed) for harmonizing MRI-derived measurements due to its ease of application and ability to adjust for site-related additive and multiplicative biases while preserving biological variability \cite{Radua2020, Fortin2018, Cai2021, Leithner2022}.
Its success has led to the development of various ComBAT-like methods, such as Longitudinal ComBAT~\cite{Beer2020}, CovBat~\cite{Chen2022}, Auto-ComBAT~\cite{Carre2022}, GMM-ComBAT~\cite{Horng2022}, M-ComBAT~\cite{Da-ano2020}, and ComBAT-GAM~\cite{Pomponio2020}, Cluster-ComBAT~\cite{Hoang2024}, each offering specific improvements. Despite these advances, the original ComBAT remains a widely used solution in both research and clinical research applications~\cite{Radua2020, Fortin2018, Marzi2024}.

ComBAT relies on several key assumptions, which researchers have highlighted in previous studies. Failing to meet these assumptions can significantly impact its effectiveness. The primary assumptions include:
\begin{enumerate}
    \item The covariate effects (i.e. sex, age, handedness, etc.) on the data must be consistent across all harmonization sites (more details in \S~\nameref{sec:theory}).
    \item The population distributions must be uniform across sites, with no substantial imbalances in key covariates (i.e., sample, age, sex, medical condition)~\cite{Bayer2022, Orlhac2022, Parekh2022}.
    \item Age distributions must overlap substantially across sites and span a wide age range~\cite{Beer2020, Pomponio2020}. 
    \item Harmonization studies are limited to a manageable number of sites, and data can be harmonized in a single step once the data acquisition is complete.
\end{enumerate}

Unfortunately, these assumptions are often overlooked in practice. Notably, no study has systematically evaluated their impact on ComBAT's performance under controlled conditions or provided clear, quantifiable guidelines in diffusion MRI.

In this paper, we thoroughly review ComBAT’s mathematical foundation, clarifying its assumptions and the demographic conditions necessary for optimal performance. We also argue that aligning each site with a common reference dataset, with a method called {\em Pairwise-ComBAT}, mitigates some of ComBAT’s limitations. Doing so enhances consistency and supporting reproducibility, two essential factors for open science, collaborative research, and real-life clinical deployment. %This approach leads to , a harmonization method designed for alignment to a reference site. %with an added goodness-of-fit measure to assess harmonization quality.

Throughout our experiments, we use the Cambridge Centre for Ageing Neuroscience dataset (CamCAN) \cite{Shafto2014,Taylor2017}, as a reference dataset and evaluate the impact of the magnitude additive and multiplicative bias, and various demographic characteristics including sample size, age distribution, missing covariates and pathological population presence. In some cases, we replicate our results with other public databases, such as the %UK Biobank \cite{Sudlow2015}, the
Alzheimer’s Disease Neuroimaging Initiative (ADNI)~\cite{Aisen2010} dataset or the Mind Research Network (MRN, site A) dataset~\cite{Gollub2013}. To assess the harmonization quality, we also proposed a goodness-of-fit measure mathematically derived from {\em Pairwise-ComBAT}.

Following these experiments, we present five essential recommendations to use ComBAT. This paper aims to highlight both the limitations and opportunities that ComBAT presents for multi-site data harmonization.

% Describe ComBAT As in Fortin
\section{Theory}
\label{sec:theory}
\begin{figure*} % "t" option places the figure at the top of the page
    \centering
    \includegraphics[width=0.99\textwidth]{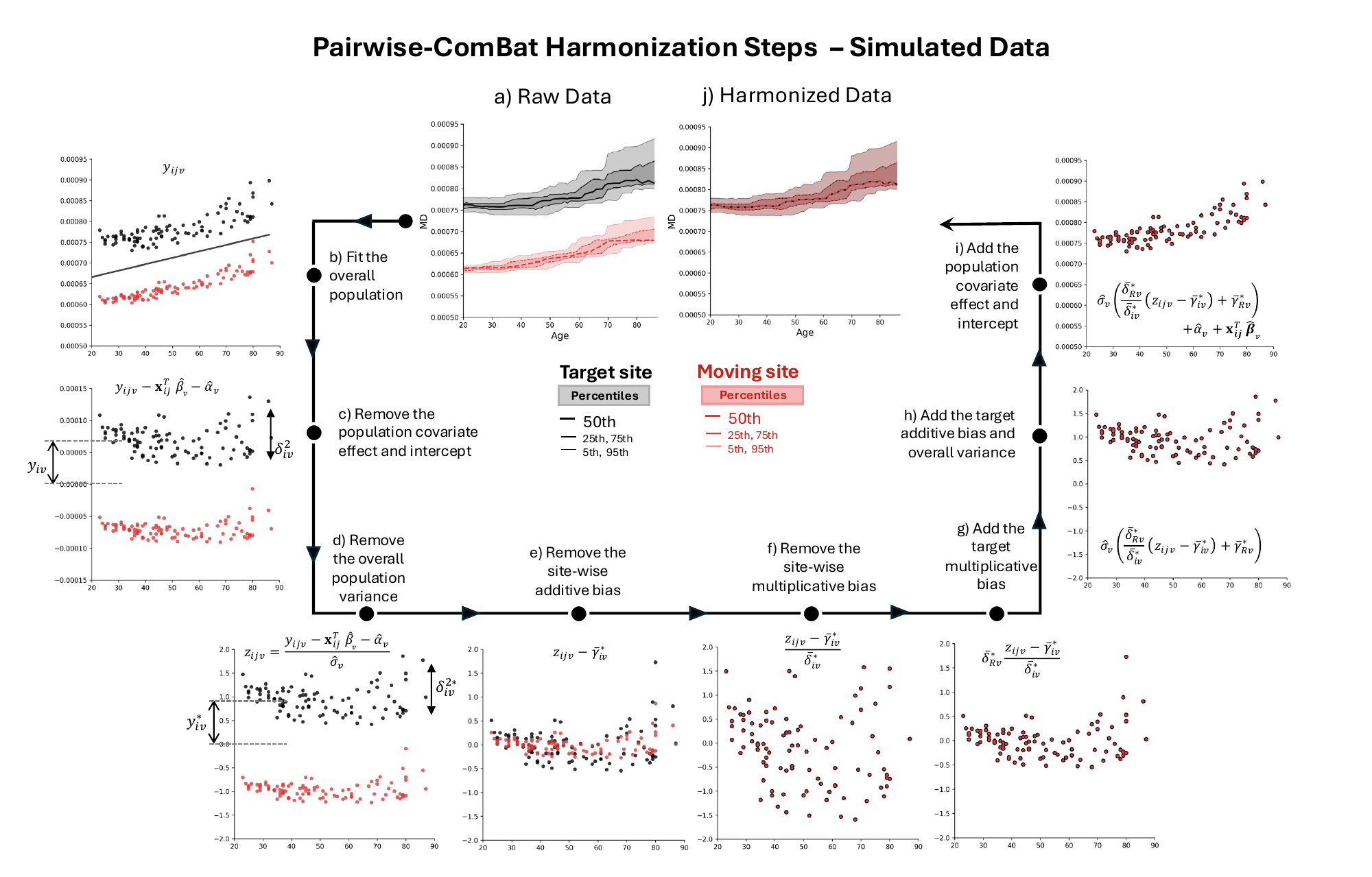}\vspace{-0.3cm}
    \caption{Illustration of the seven steps of a typical Pairwise-ComBAT harmonization of two sites underlying the effect of each variable of Eq.(17). From the raw data in a) to the harmonized data in j). The gray curves in b) illustrate the overall trend of the population from site 1 and site 2, whereas the scatter plots show the values of the $J$ subjects of each site.\vspace{-0.6cm}}
    \label{fig:step-by-step-combat}
\end{figure*}

In this section, we present the theory behind ComBAT. We outline the parameters of ComBAT, how they are estimated, and how they are used to harmonizing data. To ensure a good understanding of the method, the reader shall refer to Figure~\ref{fig:step-by-step-combat} for the key harmonization steps. Note that although this figure is tailored to Pairwise-ComBAT (c.f. \S \nameref{sec:pairwisecombat}), it can be used to demystify ComBAT's formalism.

Let $\mathcal{Y}=\{ Y_1, ..., Y_I\}$ be a set of $I$ physical MRI sites and $Y_i=\{ Y_{i1}, ..., Y_{iJ_i}\}$ the $J_i$ dMRI derived metric images from the site $i$.  %, all non-linearly registered onto a common space.  
For example, $Y_i$ may contain a set of FA maps obtained from the same MRI system at site $i$.  % and registered onto the MNI space \cite{fonov_unbiased_2009, fonov_unbiased_2011}). 
Note that $J_i$ is site dependent, as the number of data is usually different from one site to another. Each map $Y_{ij}$ can be represented as $Y_{ij}=[y_{ij1}, ..., y_{ijV}]$, where $V$ is the number of voxels (or regions) in the common space and $y_{ijv} \in \mathbb{R}$. % In this paper, $V$ is the number of brain voxels in the MNI space.   %The $V$ features can be voxels or various regions of interest, such as white matter bundles. 

To compensate for site-specific biases, ComBAT uses a linear model for the data formation of each voxel (or region) $v$ as follows:  
\begin{equation}
\label{eq:combat}
y_{ijv} = \alpha_v + \textbf{x}^T_{ij}\bm{\beta}_v + \gamma_{iv} + \delta_{iv}\varepsilon_{ijv}, 
\end{equation}
where $\alpha_v$ is the model intercept of the overall population across all sites, $\textbf{x}_{ij}$ is a vector of covariates (i.e. age and sex), $\bm{\beta}_v$ is a regression vector, $\gamma_{iv}$ is the additive site effect, $\delta_{iv}$ is the multiplicative site effect, and $\varepsilon_{ijv}$ is some random independent Gaussian noise with $\varepsilon_{ijv} \sim \mathcal{N}(0,\sigma^{2}_{v})$.  This model is illustrated for two sites in Figure~\ref{fig:step-by-step-combat}(b) where the black line corresponds to the overall population trend represented by $\beta_v$ in the previous equation. It is crucial to recognize that ComBAT assumes the regression vector $\bm{\beta}_v$ (i.e. the slope of the population) is constant for all sites which, as will be demonstrated, is often incorrect and results in flawed harmonization functions. Thus, the harmonization process of two parallel black and red populations illustrated in figure~\ref{fig:step-by-step-combat}(b) is an idealized version of the algorithm. 

The primary goal of ComBAT is to eliminate the site-specific additive and multiplicative biases (respectively $\gamma_{iv}$ and $\delta_{iv}$) ensuring that the harmonized population profile conforms to the black dotted model depicted in Figure~\ref{fig:step-by-step-combat}(j):
\begin{equation}
y_{ijv}^{ComBAT} = \alpha_v + \textbf{x}^T_{ij}\bm{\beta}_v + \varepsilon_{ijv}. \label{eq:combat_cannonical}
\end{equation}

Since each parameter of the model $\alpha_v, \bm{\beta}_v, \gamma_{iv}, \sigma_v$ and $\delta_{iv}$ are a priori unknown, they must be empirically estimated from the data.  % If these parameters can be accurately estimated (referred to as $\hat{\alpha}_v, \hat{\bm{\beta}}_v, \hat{\gamma}_{iv}$ and $\hat{\delta}_{iv}$), the harmonized values $y_{ijv}^{ComBAT}$ can be computed as follows:
%
%\begin{eqnarray}
%y_{ijv}^{ComBAT} = \frac{y_{ijv} - \hat{\alpha}_v - \textbf{x}^T_{ij}\hat{\bm{\beta}}_v - \hat{\gamma}_{iv}}{\hat{\delta}_{iv}} + \hat{\alpha}_v +\textbf{x}^T_{ij}\hat{\bm{\beta}}_v.
%\end{eqnarray}
%
One simplistic approach to estimating these parameters is through the L/S-ComBAT (Location and Scale) method~\cite{Johnson2007}. Although this approach is straightforward, it has its share of limitations which are typically addressed with the Bayesian formulation of ComBAT which is simply called {\em ComBAT} in Fortin et al. (2017) \cite{Fortin2017}.

ComBAT starts by computing the regression vector $\bm{\beta}_v$ and the intercept $\alpha_v$ with an ordinary least-square approach
\begin{equation}
[\hat \alpha_v,\hat \beta_v] = (X^T X)^{-1} X^T Y_v,  
\end{equation}
where $X$ is the matrix of covariate vectors from every participant across all sites, and $Y_v$ is a vector containing the metric in location $v$ for every participant in all sites.

ComBAT standardizes all data in relation to $\hat \alpha_v$ and $\hat \sigma_v$ as depicted in Figure~\ref{fig:step-by-step-combat}(d) and represented mathematically as 
\begin{equation}
z_{ijv} = \frac{y_{ijv} - \hat{\alpha}_v - \textbf{x}^T_{ij}\hat{\bm{\beta}}_v}{\hat{\sigma}_v},
\label{eq:zijv}
\end{equation}
where $\hat \sigma_v$ is given by 
\begin{equation}
\hat{\sigma}^2_v =\frac{1}{J_I} \sum_{ij}(y_{ijv} - \hat{\alpha}_v - \textbf{x}^T_{ij}\hat{\boldsymbol{\beta}}_v - \hat{\gamma}_{iv})^2,
\end{equation}
where $J_I$ is the total number of data across every site.
Then, ComBAT estimates $\gamma_{iv}^*$ and $\delta_{iv}^{2*}$  by maximizing their posterior distribution instead of their likelihood distribution as in L/S ComBAT\cite{Johnson2007}\footnote{ComBAT notation can be confusing.  While $\gamma_{iv}$ and $\delta_{iv}^{2}$ are the bias and variance of the rectified population (fig.~\ref{fig:step-by-step-combat}(c)), $\gamma_{iv}^*$ and $\delta_{iv}^{2*}$ are bias and variance of the standardized populations of fig.~\ref{fig:step-by-step-combat}(d)  }.  Following Bayes' theorem, the posteriors $\gamma^*_{iv}$ and $\delta_{iv}^{2*}$ can be written as 
\begin{align}
\pi(\gamma^*_{iv} | \textbf{z}_{iv}, \delta_{iv}^{2*}) &\propto L(\textbf{z}_{iv}|\gamma^*_{iv}, \delta_{iv}^{2*})P(\gamma^*_{iv}), \label{eq:posteriors1}\\
\pi(\delta^{2*}_{iv} | \textbf{z}_{iv}, \gamma^*_{iv}) &\propto L(\textbf{z}_{iv}|\gamma^*_{iv}, \delta_{iv}^{2*})P(\delta^{2*}_{iv}),\label{eq:posteriors2}
\end{align}
where 
\begin{align*}
 L(\textbf{z}_{iv}|\gamma^*_{iv}, \delta_{iv}^{2*}) &= \mathcal{N}(\gamma^*_{iv}, \delta_{iv}^{2*}) \mbox{ (likelihood - Gaussian)}\\
 P(\gamma^*_{iv}) &= \mathcal{N}(\mu_{i}, \tau_{i}^2) \mbox{  (prior - Gaussian)} \\
 P(\delta^2_{iv}) &= \mathcal{IG}(\lambda_{i}, \theta_{i}) \mbox{  (prior - Inverse Gamma)}. 
\end{align*}
The hyperparameters of the prior distributions $\mu_i$, $\tau^2_i$, $\lambda_i,$ and $\theta_i$ are estimated following the moment of these distributions.  As such, $\mu_i$ and $\tau^2_i$ are the empirical mean and variance across every location $v$ of site $i$,
\begin{align}
\bar{\mu}_{i} = \frac{1}{V}\sum_v \hat{\gamma}^*_{iv}, \;\;\;\;\;
\bar{\tau}^2_{i} = \frac{1}{V-1}\sum_v (\hat{\gamma}^*_{iv} - \bar{\mu}_{i})^2, \label{eq:moments_gamma}
\end{align}
where $\hat{\gamma}^*_{iv} = \frac{1}{J_i}\sum_j z_{ijv}$ is the location-wise and site-wise average of the \textbf{standardized} data illustrated in figure~\ref{fig:step-by-step-combat}(d) and (e). 

The estimation of $\lambda_i,$ and $\theta_i$, one first need to compute the voxel-wise and site-wise standardized variance $\hat{\delta}^{2*}_{iv} = \frac{1}{J_i-1}\sum_j (z_{ijv} - \hat{\gamma}^*_{iv})^2$.   The empirical mean and variance of $\hat{\delta}^{2*}_{iv}$ across all location $v$ is computed as $\bar{G}_{i} = \frac{1}{V}\sum_v \hat{\delta}^{2*}_{iv}$ and $\bar{S}^{2}_{i} = \frac{1}{V-1}\sum_v (\hat{\delta}^{2*}_{iv} - \bar{G}_{i})^2$.  Then, by having $\bar{G}_i$ and $\bar{S}^2_i$ equal to the first and second theoretical moment of the inverse gamma distribution, we get 
\begin{equation}
\bar{G}_{i} = \frac{\theta_i}{\lambda_i - 1},\;\;\;\;\;\; %\label{eq:moments_delta_theoritical1},\\
\bar{S}^{2}_i = \frac{\theta_i^2}{(\lambda_i - 1)^2(\lambda_i - 2)} \label{eq:moments_delta_theoritical2}
\end{equation}
that can be rearranged as follows to get the estimate of the two hyperparameters
\begin{equation}
\bar{\lambda}_i = \frac{\bar{G}_i^{2} + 2\bar{S}^{2}_i}{\bar{S}^{2}_i}, \;\;\;\;\;
\bar{\theta}_i = \frac{\bar{G}_i^{3} + \bar{G}\bar{S}^{2}_i}{\bar{S}^{2}_i}. \label{eq:lambda_theta}
\end{equation}
Now that the hyperparameters of the likelihood and prior distributions have been estimated, by developing eq.(\ref{eq:posteriors1}) and (\ref{eq:posteriors2}) and combining to it eq.~(\ref{eq:moments_gamma}) and (\ref{eq:lambda_theta}) to it, we get that the mathematical expectation of the posterior distributions lead to the following estimate of $\gamma^*_{iv}$ and $\delta^{2*}_{iv}$   
\begin{eqnarray}
\hat{\gamma}^*_{iv} = \hat{\mathbb{E}}(\gamma^*_{iv}| \textbf{z}_{iv}, \delta_{iv}^{2*}) = \frac{J_i \bar{\tau}_i^2\hat{\gamma}^*_{iv} + \bar{\delta}_{iv}^{2*}\bar{\mu}_i}{J_i \bar{\tau}_i^2 + \bar{\delta}_{iv}^{2*}} \label{eq:gamma_star}\\
\hat{\delta}_{iv}^{2*} = \hat{\mathbb{E}}(\delta^{2*}_{iv}| \textbf{z}_{iv}, \gamma_{iv}^{*}) = \frac{\bar{\theta}_i + \frac{1}{2}\sum_j (z_{ijv} - \bar{\gamma}_{iv}^{*})^2}{\frac{J_i}{2} + \bar{\lambda}_i - 1}.\label{eq:delta_star}
\end{eqnarray}

Since equations (\ref{eq:gamma_star}) and (\ref{eq:delta_star}) are inter-dependent, $\bar{\gamma}_{iv}^*$ and $\bar{\delta}_{iv}^{2*}$ are estimated through an iterative process by starting with a reasonable value for $\bar{\delta}_{iv}^{2*}$ (i.e., $\hat{\delta}^{2*}_{iv}$), than by calculating $\bar{\gamma}_{iv}^*$, and then by re-estimating $\bar{\delta}_{iv}^{2*}$ with the new value of $\bar{\gamma}_{iv}^*$ and so on.  This process is repeated until converges. 

Once all parameters of eq.(\ref{eq:combat}) have been empirically estimated, the data can be harmonized as follows :

\begin{equation}
y_{ijv}^{ComBAT} = \frac{\hat{\sigma}_v}{\bar{\delta}_{iv}^*}(z_{ijv} - \bar{\gamma}_{iv}^*) + \hat{\alpha}_v + \textbf{x}^T_{ij}\hat{\bm{\beta}}_v,
\label{eq:apply}
\end{equation}
as illustrated in Figure~\ref{fig:step-by-step-combat}(f) to (i).

\subsection{Mathematical limitations of ComBAT}
\label{sec:limitations}

\paragraph{The slope $\bm \beta_v$} As outlined through the previous equations, ComBAT assumes that the slope $\beta_v$ is the same for all sites' region $v$.  This assumption is problematic when the canonical data formation model of Eq.(\ref{eq:combat_cannonical}) is affected by a multiplicative factor $S_i$ and an additive factor $A_i$ as follows: 
\begin{eqnarray}
\label{eq:combat_M_A_Bias}
y_{ijv} &=& y_{ijv}^{ComBAT}*S_i + A_i \nonumber \\
&=& \alpha_vS_i + A_i + \bm{x}_{ij}^T\bm{\beta}_vS_i + \epsilon_{ijv}S_i \nonumber \\
&=& \alpha_v + \gamma_{iv} + \bm{x}_{ij}^T\bm{\beta}_vS_i + \delta_{iv}\epsilon_{ijv},
\end{eqnarray}
where $\gamma_{iv}=\alpha_v(S_i-1)+A_i$ and $\delta_{iv}=S_i$. While Eq.~(\ref{eq:combat_M_A_Bias}) is similar to Eq.~(\ref{eq:combat}), it nonetheless contains a multiplicative term $S_i$ on the slope which is not taken care of by ComBAT. As will be shown later, this results into misalignment populations when the uniform slope assumption is violated.  

\paragraph{The variance $\delta_{iv}^2$}
When calculating the variance $\hat{\delta}_{iv}^{2*}$, two key considerations ought to be taken into account. First, the posterior estimator of Eq.~(\ref{eq:delta_star}) can be viewed as a nonlinear weighted average of the empirical and prior variances with respect to the $J_i$ (i.e. the number of data at site $i$). To illustrate that, when $J_i \rightarrow 0$,  Eq.~(\ref{eq:delta_star}) becomes $\hat{\delta}_{iv}^{2*} \rightarrow \frac{\bar \theta_i}{\bar \lambda_i - 1}$, which, in combination with Eq.~(\ref{eq:lambda_theta}), yields to $\hat{\delta}_{iv}^{2*}  \approx \hat G_i$, representing the prior variance. Conversely, as $J_i \rightarrow \infty$, Eq.~(\ref{eq:delta_star}) simplifies to $\hat{\delta}_{iv}^{2*} \rightarrow \frac{1}{J_i} \sum_j (z_{ijk} - \gamma_{iv})^2$, the formula for empirical variance. However, as we will demonstrate empirically, the weight assigned to the prior term remains disproportionately high as $J_i$ increases.  As a consequence, a poor a priori variance $\hat G_i$ will affect $\hat{\delta}_{iv}^{2*}$ even on large populations of more than $\sim\!\! 100$ subjects.

The second consideration that one must keep in mind is the fact that the variance $\hat{\delta}_{iv}^{2*}$  depends on $z_{ijv}$. Since $z_{ijv}$ is itself influenced by $x_{ij}^T \bm \beta_v$ (c.f. Eq.~(\ref{eq:zijv})), whenever the slope assumption is violated, the term $(z_{ijv} - \bar \gamma_{iv}^*)^2$ becomes biased, leading to an inaccurate estimation.

%%%%%%%%%%%%%%%%%%%%%%%%%%%%%%%%%%%%%%%%%%%%%%%%%%%%%%%%%%%%%%%%%%%%%%%%%%%%%%%%%%%%%%%%%%%%%%%%
\section{Method}
\subsection{For the need of a reference site}

ComBAT harmonizes multisite data by projecting each data point onto an "average" site, defined by an intercept $\hat{\alpha}_v$, a slope $\hat{\bm \beta}_v$, and a variance $\hat{\sigma}_v^2$ (Eq. (\ref{eq:apply})), illustrated by the black line in Fig.~\ref{fig:step-by-step-combat}(b). This approach is ideal for structured studies or clinical trials, in which participants undergo specific interventions according to a predefined protocol, with data collected at multiple sites under controlled conditions. Additionally, such studies are usually time-bound, culminating with a fixed set of sites and participants that are harmonized collectively at the end of the study.

However, many applications do not meet these constraints. We argue that aligning each site to a common reference dataset, ideally one that is well populated across all covariates, offers a more versatile and robust solution. There are three main reasons for adopting this approach.

\vspace{-0.2cm}

\paragraph{\textbf{i) Reproducibility and open science}} Harmonizing data to a reference dataset offers multiple advantages for the scientific community. First, it allows for the creation of openly accessible and readily usable datasets, as it enhances the comparability of data collected from diverse sources. This eliminates the need for future re-harmonization and simplifies data sharing processes. By aligning with a common reference site, researchers can also establish reproducible standards for critical imaging metrics, such as those in diffusion MRI, which play a key role in characterizing disease. This reproducibility is essential to open science, fostering transparent and collaborative research across institutions. Additionally, harmonization to a reference site facilitates longitudinal studies, where data collection spans extended periods. When new data are harmonized to the same reference, researchers can confidently track disease progression or treatment outcomes without concerns about changing the harmonization parameters. This stability supports the robustness of findings over time and reduces the computational resources needed for repeated harmonization.

\vspace{-0.2cm}
\paragraph{\textbf{ii) Normative modeling}} Normative modeling quantifies individual variability by assessing how much a subject deviates from a representative population. Typically, a normative model is built by pooling data across sites. Since ComBAT's parameters ($\alpha_v, \bm \beta_v$, and $\sigma_v$) are computed from all sites’ data, adding new data to the pool shifts the average reference site, creating a "moving target" issue. Although Kia et al. propose a hierarchical Bayesian regression (HBR) approach to mitigate this effect \cite{Kia2022}, this approach cannot be used to harmonize data. Also, we suggest that harmonizing each site pairwise to a stable reference dataset is a much simpler approach for building a normative reference model than the iterative and cyclic approach that HBR is built upon.

\vspace{-0.2cm}

\paragraph{\textbf{iii) Routine clinical practice}} Clinical practice entails a continuous flow of data from numerous, and potentially expanding, sources over an indefinite period. For software providers, this results in the need to accommodate unpredictable growth in data sources. This variability challenges the traditional ComBAT harmonization method. By independently harmonizing each site to a common reference, new sites can be seamlessly integrated as they come online, avoiding the common ComBAT issue where the harmonization of a new site disrupts the parameters already established for previously harmonized sites. Furthermore, once harmonization parameters are set for a site, they can be applied to incoming data from that site, thus reducing the need for repeated harmonization efforts.

In our experiments, we examine how limited data at either the target or reference site, restricted age ranges, and undocumented diseases affect ComBAT's performance. Unlike previous studies that focus solely on training error, we treat harmonization as a machine learning problem with distinct training and testing sets. This approach enables us to evaluate ComBAT’s generalization to new, unseen data, providing a more comprehensive assessment of its effectiveness.

\subsection{Pairwise-ComBAT}
\label{sec:pairwisecombat}

The goal of {\em Pairwise-ComBAT} is to harmonize each moving site, denoted as $M$, independently onto a reference site, $R$. The reference site acts as a standardized baseline, containing diverse data representative of healthy individuals.

To achieve this, data from site $M$ is harmonized using the parameters of site $R$, specifically the multiplicative ($\bar{\delta}_{Rv}^*$) and additive ($\bar{\gamma}_{Rv}^*$) batch effects. This results in a modified version of Eq. (\ref{eq:apply}):
\begin{equation}
y_{Mjv}^{ComBAT} = \hat{\sigma}_v \left( \frac{\bar{\delta}_{Rv}^*}{\bar{\delta}_{Mv}^*} (z_{Mjv} - \bar{\gamma}_{Mv}^*) + \bar{\gamma}_{Rv}^* \right) + \hat{\alpha}_v + \mathbf{x}_{Rj}^T \hat{\bm{\beta}}_v.
\label{eq:pairwisecombat}
\end{equation}
For a more intuitive understanding of Pairwise-ComBAT, the reader may refer to Figure~\ref{fig:step-by-step-combat}, particularly panels (f) to (i), which illustrate the effect of each parameter. 

While Pairwise-ComBAT is conceptually similar to Da-Ano et al's M-ComBAT~\cite{Da-ano2020}, our harmonization equation (Eq. (\ref{eq:pairwisecombat})) differs from their's (Eq. (4) in paper ~\cite{Da-ano2020}). Furthermore, in all our experiments, we assume the reference site is well-populated, an assumption not made by Da-ano et al.

The reader shall note that Pairwise-ComBAT is only a reconfiguration of the original ComBAT method, thus producing an equivalent harmonization outcome with the same strengths and weaknesses as the original ComBAT~\cite{Fortin2017}.

\subsection{ Goodness of fit}

To gauge the performances of ComBAT, we propose a goodness-of-fit metric that measures how well two populations overlap. By assuming that the moving (M) and reference (R) populations are well harmonized, our method rectifies both populations using the reference parameter intercept $\bar \gamma_{Rv}^*$, i.e. :
\begin{eqnarray}
z_{Rjv} &=& y_{Rjv}-\hat \alpha_v - \bar \gamma_{Rv}^* - \mathbf{x}_{Rj}^T \beta_v    \;\;\;\;\;\;\;\;\;        \forall y_{Rjv} \in T_{Tv}      \nonumber \\                            
z_{Mjv}&=& y_{Mjv}^{\mbox{\tiny ComBAT}}-\hat \alpha_v - \bar \gamma_{Rv}^* - \mathbf{x}_{Mj}^T \beta_v      \;\;\;   \forall y_{Mjv} \in Y_{Mv}                  \nonumber             
\end{eqnarray}
By removing the effect of the covariates and the biases, this rectification allows assuming that $z_{ijv}$ is independent from $\mathbf{x}_{ij}$ and that $P(z_{Rjv} |\mathbf{x}_{Rj})=P(z_{Rjv})$ and $P(z_{Mjv} |\mathbf{x}_{Mj})=P(z_{Mjv})$ which is Gaussian as mentioned after eq.~(\ref{eq:posteriors2}).
The distance between univariate distribution is computed using the Bhattacharyya distance (BD) which, in the case of two Gaussian distributions, results into the following closed-form solution
\begin{equation}
d(P(z_{Rjv})|P(z_{Mjv})) = \frac{1}{4} \frac{(\mu_{Tv}-\mu_{Mv})^2}{(\sigma_{Rv}^2+\sigma_{Mv}^2)} +  \frac{1}{2}\ln((\frac{\sigma_{Tv}^2+\sigma_{Mv}^2}{2\sigma_{Tv} + \sigma_{Mv}}) \nonumber
\end{equation}
where $\mu_{Rv}=mean_j(z_{Rjv})$, $\mu_{Mv}=mean_j(z_{Mjv})$, $\sigma_{Rv}^2=var_j(z_{Rjv})$, and $\sigma_{Mv}^2=var_j(z_{Mjv})$.

\begin{figure*}[t] % "t" option places the figure at the top of the page
    \centering
    \includegraphics[width=.8\linewidth]{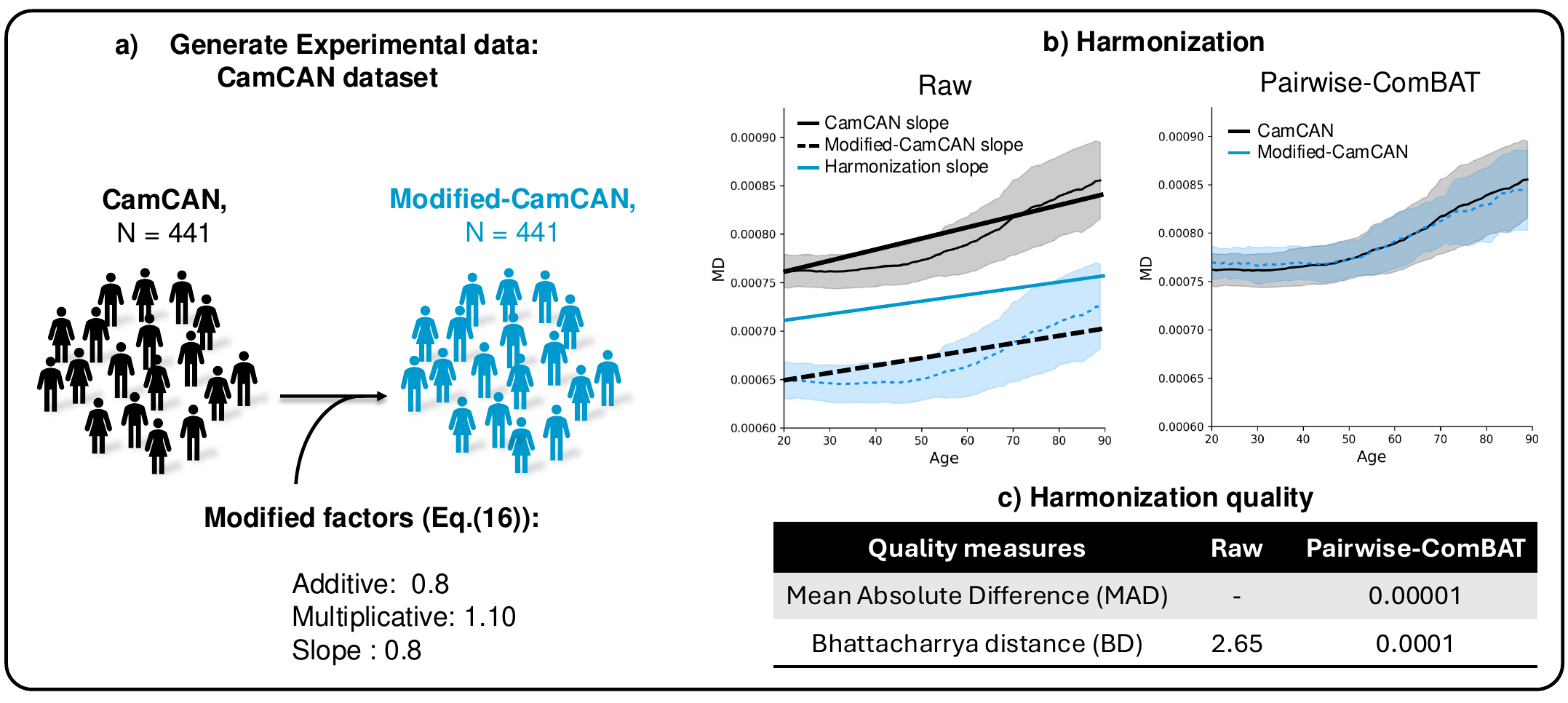}
    \caption{Pairwise-ComBAT harmonization of the mean diffusivity (MD) in the {\em modified CamCAN} dataset against the unbiased CamCAN (N=441). (a) The parameters A (additive), M (multiplicative), S (slope) used to generate the modified CamCAN version (c.f. Eq.(\ref{eq:combat_M1_M2_A_Bias}). (b) (left) raw data with the slopes for CamCAN (solid black line), modified CamCAN (dashed black line), and the slope estimated by Pairwise-ComBAT (solid blue line). (right) The Pairwise-ComBAT harmonization. (c) Distances between the raw and harmonized populations.}
    \label{fig:CamCAN-vs-itself}
\end{figure*}

An illustration of the BD computed before and after harmonization is provided in Figure~\ref{fig:CamCAN-vs-itself}.

\subsection{Dataset studied}

\subsection*{Reference site}

The public CamCAN Stage 2 dataset \cite{Shafto2014,Taylor2017} is a large-scale, multi-modal study investigating cognitive aging across the adult lifespan (ages $18-87$) using MRI, MEG, and cognitive data. With roughly 700 participants, the sample is well-balanced, thanks to a uniform age and sex distribution and recruitment from a larger population-based cohort of 2681 individuals \cite{Taylor2017}. Approval for the study was granted by the Research Ethics Committee of Cambridgeshire 2 (reference: 10/H0308/50 \cite{Shafto2014}), and participants gave written, informed consent prior to involvement. A total of 441 healthy participants (224 males, 217 females) were selected for this study. 

DWI data were acquired on a 3T Siemens TIM Trio Scanner (32-channel head coil) using a twice refocused spin echo (TRSE) echo-planar imaging sequence to reduce eddy-current artifacts. The acquisition included two b-values of $1000$ and $2000\,\mathrm{s}/\mathrm{mm}^2$ along 30 directions and three non-diffusion-weighted images (b=0). Additional parameters were: 66 axial slices, voxel size = 2 mm x 2 mm x 2 mm, repetition time (TR) = 9100 ms, echo time (TE) = 104 ms, and an acceleration factor of 2 using GRAPPA \cite{Taylor2017}. 

\subsection*{Moving sites}

\paragraph{Modified-CamCAN}

This dataset contains the same data as the CamCAN dataset but with different additive and multiplicative biases following this equation:
\begin{equation}
\label{eq:combat_M1_M2_A_Bias}
y_{ijv} = \alpha_v + \gamma_{iv}A + \bm{x}_{ij}^T\bm{\beta}_vS + \delta_{iv}\epsilon_{ijv}M
\end{equation}
where $A$ is used to translate the population, $S$ to change the slope of the population wrt their associated covariates and $M$ to change the variance of the population. 

An example of CamCAN and modified CamCAN (both with 441 subjects) with $A=0.8$, $S=0.8$ and $M=1.1$ is illustrated in Figure~\ref{fig:CamCAN-vs-itself}(a).  Figure~\ref{fig:CamCAN-vs-itself}(b) shows the two population distributions before (left) and after (right) the Pairwise-ComBAT harmonization. In Figure~\ref{fig:CamCAN-vs-itself}(c), we have the BD metrics and mean absolute difference (MAD) between the moving and the reference site data points before and after harmonization. This figure echoes with most experiments reported in \S~\nameref{sec:results}.

\paragraph{ADNI}

The ADNI data were obtained from the Alzheimer’s Disease Neuroimaging Initiative (ADNI) database\footnote{adni.loni.usc.edu}. The ADNI was launched in 2003 as a public-private partnership, led by Principal Investigator Michael W. Weiner,MD. The primary goal of ADNI has been to test whether serial magnetic resonance imaging
(MRI), positron emission tomography (PET), other biological markers, and clinical and neuropsychological assessment can be combined to measure the progression of mild cognitive impairment (MCI) and early Alzheimer’s disease (AD).

We selected the 152 subjects (100 Healthy Control (HC), 28 Mild Cognitive Impairment (MCI), 24 Alzheimer's Disease (AD)) from site 127-GE site of the ADNI-3 cohort \cite{Aisen2010}. ADNI-127-GE includes $83$ males and $69$ females, with age mean/std of $73.4 \pm 7.4$ (Min: 61, Max: 90), $142$ subjects were left-handed and 11 right-handed.

The diffusion weighted MRI (DWI) data were acquired on a 3T General Electric (GE) Healthcare using a spin echo imaging sequence. Acquisition includes 6 non-diffusion-weighted images (b=0) and 48 directions with b-value of $1,000\,\mathrm{s}/\mathrm{mm}^2$. Additional parameters are: TE =56 ms, TR =7800 ms, voxel size = 2 mm x 2 mm x 2 mm and approximate scan time =7 min 30 s.

\paragraph{NIMH}
We used the 157 subjects from the National Institute of Mental Health (NIMH) Intramural Healthy Volunteer Dataset~\cite{Nugent2024data}. 
The MRI protocol used was initially based on the ADNI-3 basic protocol and was modified for DWI acquisition to include the slice-select gradient reversal method and reversed blip scans; and turned off reconstruction interpolation. The 3D-T1 weighted acquisition from ADNI-3 
was replaced by the 3D-T1 acquisition from the ABCD protocol.
All participants provided electronic informed consent for online pre-screening, and written informed consent for all other procedures.

\subsection{Processing}
\label{subsec:experiment-data}

The same tractoflow pipeline \cite{Theaud2020} was used to process the datasets.  b-values below $b=1200\,\mathrm{ms}/\mathrm{\mu m}^2$ were used to compute DTI-derivatives scalar maps and b-values above $b=700\,\mathrm{ms}/\mathrm{\mu m}^2$ were used to compute fiber orientation distribution function (fODF) metrics scalar maps. The fODF was generated using a spherical harmonics order of $8$ and the same fiber response function~\cite{Descoteaux2008}  for all the subjects ($15, 4, 4$) x $10 ^4\mathrm{ms}/\mathrm{\mu m}^2$. Next, all metrics were registered to the Montreal Neurological Institute (MNI) space. The IIT Human Brain Atlas (IIT-FA-skeleton, v.5.0, ~\cite{Qi2021-IIT}) was finally used to extract the averaged value for each major white matter fiber bundle of interest (N=25). 

For simplicity and ease of reading, the results section presents only one white matter bundle, the arcuate fascicles left (AF left), and a single diffusion metric derived from the Diffusion Tensor Imaging (DTI) model, the mean diffusivity (MD).

\subsection{Harmonization of the experimental data}
\label{subsec:Combat-exp}

As mentioned before, we consider ComBAT as a machine learning algorithm. As such, every dataset was divided into 2 subsets: a training subset used to estimate the harmonization parameters $\hat\sigma_v, \bar\delta_{iv}^*, \bar\gamma_{iv}^*, \hat\alpha_v$ and $\bm{\hat\beta}_v$, and a test subset to gauge the generalization capabilities of ComBAT. In all cases, CamCAN was used as the reference site towards which each dataset is harmonized to. In the case of modified-CamCAN, 100 random samples of CamCAN are used for the reference site and the remaining 341 samples are transformed following Eq.(\ref{eq:combat_M1_M2_A_Bias}) into modified-CamCAN. For the other datasets, the reference site of CamCAN is made of all 441 data samples.  
Each experiment is repeated randomly 30 times for each criterion and averaged.

\begin{figure*}[h] % "t" option places the figure at the top of the page
    \centering
    \includegraphics[width=\textwidth] {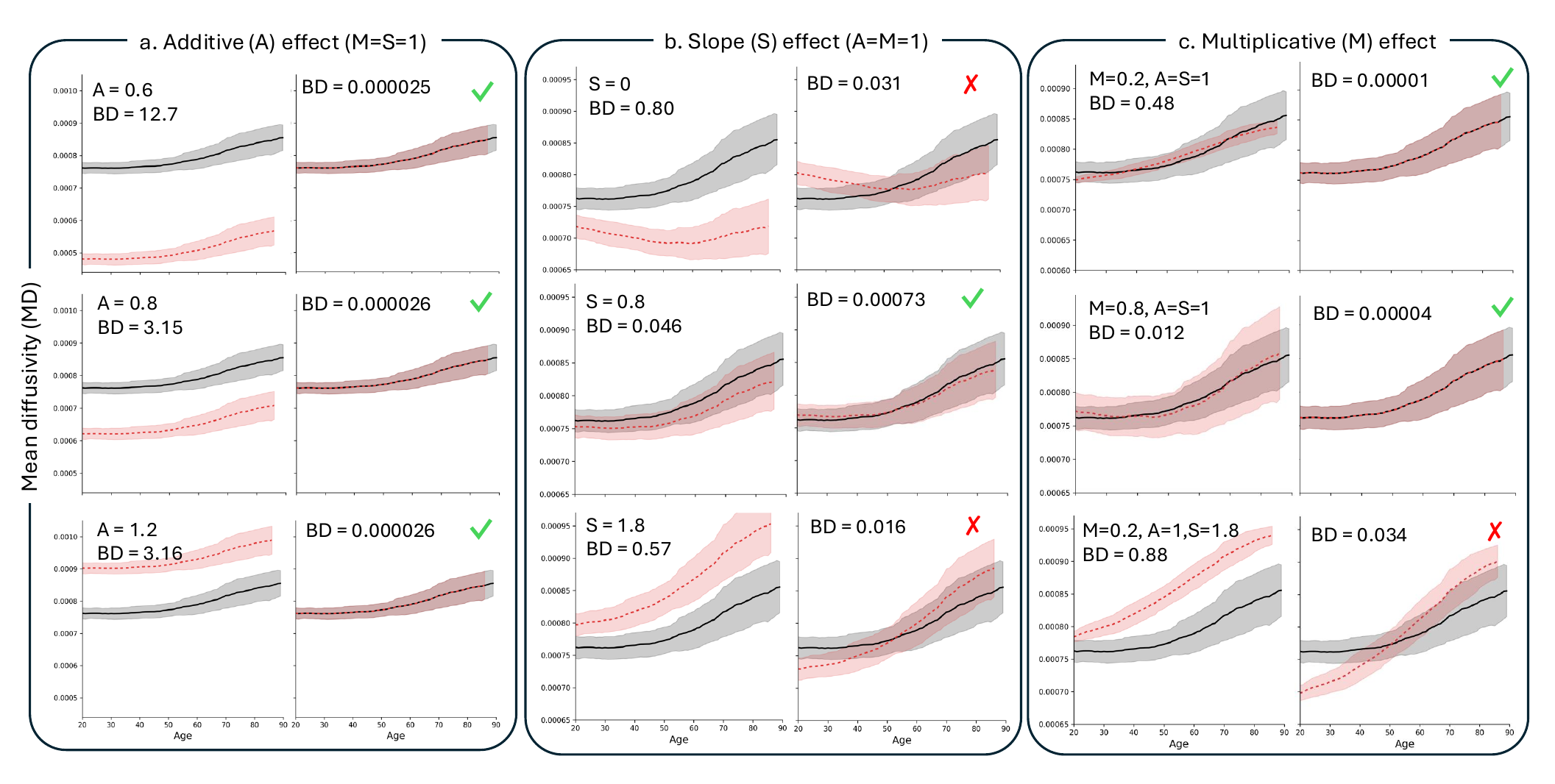}
    \vspace{-0.4cm}
    \caption{Experiment 1.  Harmonization of the Modified-CamCAN population (in gray) on the CamCAN population (in red).  Results for different multiplicative factors on (a) the bias (A) (b) the Slope (S) and (c) the noise variance (M) as described in eq.~(\ref{eq:combat_M1_M2_A_Bias}).  The left columns are the original data and the right columns are the harmonized data. Green checks indicate correct harmonization, red X's indicate poor harmonization.}
    \label{fig:factors_exp_camcan}
\end{figure*}

\subsection{Experiments}
\label{subsec:experiments}

Five experiments were conducted to gauge the performances of ComBAT when pushed to its limits. In the case of modified-CamCAN, since it contains the same data samples as CamCAN (up to a transformation - Eq.(~\ref{eq:combat_M1_M2_A_Bias})), we report the mean absolute difference (MAD) for both the training and the testing data. As for the other datasets' harmonization, we report the Bhattacharyya goodness-of-fit distance (BD).

\paragraph{Experiment 1: Additive and multiplicative biases}

 The aim of the first experiment is to illustrate that, as mentioned in \S~\nameref{sec:limitations}, ComBAT cannot correctly align two populations and properly estimate the variance of the moving site $\hat \delta_{iv}^{2*}$ when the two populations have different covariate slopes. As such, these experiments gauge the effect that the bias, slope and variance multiplicative factors have on ComBAT (respectively, $A,S$ and $M$ in Eq.~(\ref{eq:combat_M1_M2_A_Bias})).  %Train and testing error are reported with Modified-ComBAT dataset.

\paragraph{Experiment 2: Training sample size}
The effect of the number of training subjects $N$ from the moving site is evaluated by varying its value from $2$ to a maximum number of subjects. In each case, the $N$ samples are randomly selected while maintaining an equal number of males and females. In all cases, we report the training and testing MAD error obtained on the modified-CamCAN dataset.
%and provide further illustrative results on the UKBiobank dataset.

\paragraph{Experiment 3: Training age range}
The effect of the training age range is evaluated by varying the range in which the moving site data are sampled. We tested age spans from $10$ years to $60$ years, and, for each age span, we tested various age groups i.e.:

\begin{itemize}
\vspace{-0.3cm}
\item $10$ years age span : $20-30$, $30-40$, ..., $70-80$\vspace{-0.0cm}  
\item $20$ years age span: $20-40$, $30-50$, ..., $70-90$ \vspace{-0.0cm}
\item $30$ years age span: $20-50$, $30-60$, ..., $60-90$ \vspace{-0.0cm}
\item ...\vspace{-0.0cm}
\item $60$ years span: $20-80$, $30-90$.\vspace{-0.0cm}
\item $70$ years span: $20-90$.
\end{itemize}

For modified-CamCAN, samples of $32$ subjects were used for each age range, which is the maximum number of subjects possible to achieve a male/female balance with CamCAN data. Note that the $80-90$ age range was not included due to the small number of subjects.

\begin{figure}[h] % "t" option places the figure at the top of the page
    \centering
    \includegraphics[width=\linewidth]{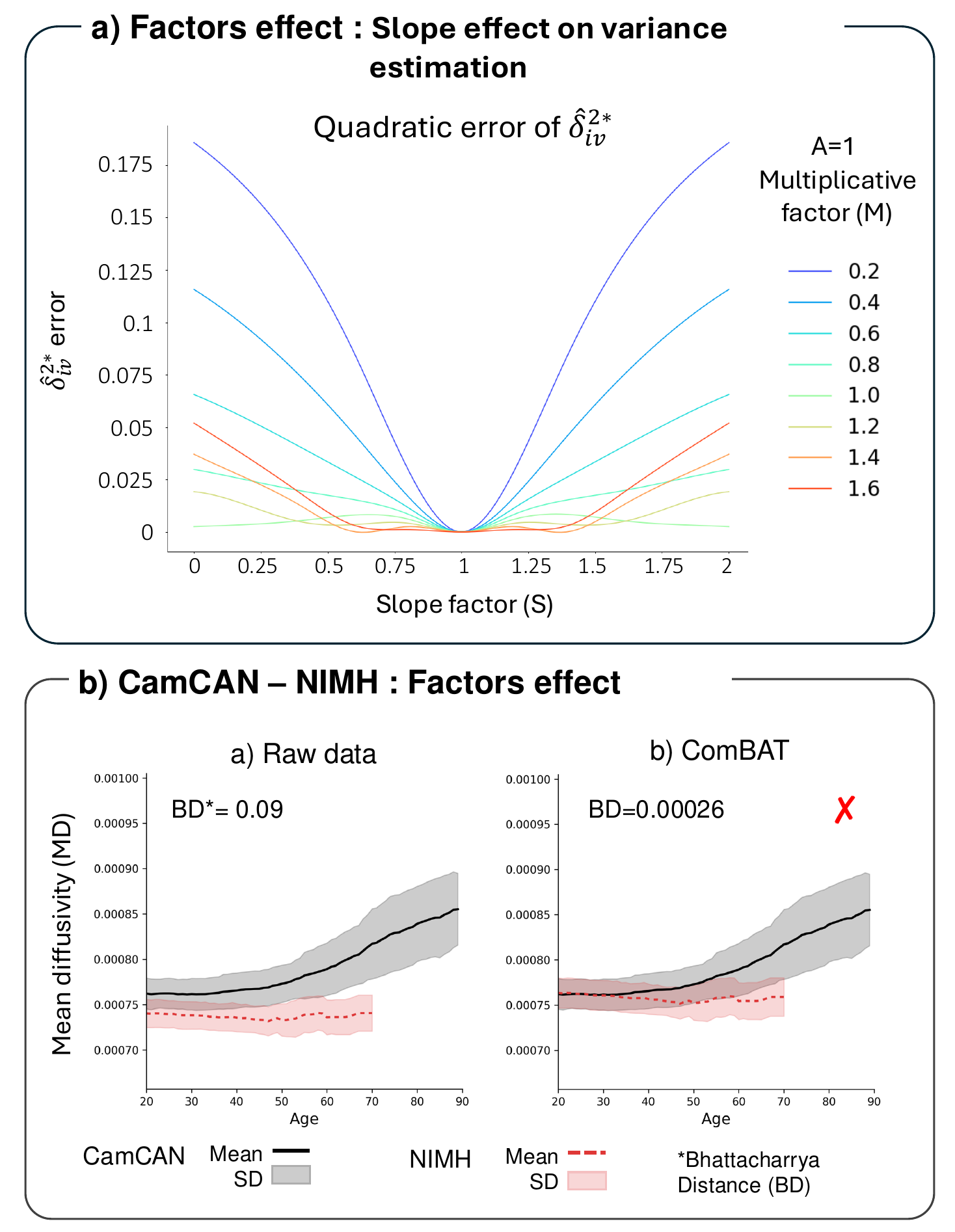}
    \caption{a) The quadratic error in the estimation of the moving site variance  $\hat \delta_{iv}^{2*}$ for different slope and variance multiplicative factors $S$ and $M$. b) Harmonization of the mean diffusivity (MD) of the NIMH population (red) onto that of CamCAN (gray). The different slopes between NIMH and CamCAN (left) leads to erroneous harmonization outcomes (right). Red cross indicates poor harmonization. \vspace{-0.3cm}% [Bottom] Mean absolute difference between the harmonized points of Modified-CamCAN and CamCAM datasets. 
    }
    \label{fig:factors_exp_var-error}
\end{figure}

\paragraph{Experiment 4: Sex covariate}
When harmonizing two populations, two key covariates are typically considered: age and sex. In this experiment, we specifically evaluated the impact of the sex covariate. The objective is twofold: first, to assess how critical the inclusion of this covariate is in the ComBAT harmonization equations, and second, to understand the importance of maintaining a well-balanced representation of males and females in the populations being harmonized.

To achieve this, we harmonized the modified-CamCAN dataset onto the CamCAN dataset under different conditions. These included scenarios with and without the sex covariate incorporated into the ComBAT equations (typically in variables $x_{ij}$ and $\beta_v$). Furthermore, we investigated the effect of excluding one of the sexes (i.e., harmonizing populations that are either male-only or balanced). This analysis allows us to measure the influence of both the inclusion of the sex covariate and the population's gender composition on the harmonization performance.

\paragraph{Experiment 5: Pathological populations}
The effect of pathology was assessed by generating a "control" group and a "pathological" group from the modified CamCAN training sample. The control group includes $100$ subjects, which remain constant throughout the experiment. The pathological group also includes 100 subjects and is modified similarly to the first experiment. The additive and multiplicative $A$ and $M$ factors of Eq.~(\ref{eq:combat_M1_M2_A_Bias}) vary from $0.8$ to $1.20$ of the initial value to simulate a more or less different distribution from the control subjects. For this experiment, the age range of the subjects are from $20$ and $90$ years, with a $50/50$ male/female distribution. The effect of pathology is assessed by calculating the harmonizer training and testing error from the healthy control group only (HC-only) as well as for the two healthy control and pathology groups (HC-patho).

\paragraph{Complementary experiments}

As previously mentioned, for readability, the following results focus on a single metric—mean diffusivity (MD)—and one white matter region—the left arcuate fasciculus (AF left). However, the same experiment was conducted on various other diffusion MRI metrics and white matter regions, all exhibiting trends consistent with those reported in the paper. The supplementary material provides results for these additional experiments, including other metrics—fractional anisotropy (FA), isotropic volume fraction (isoVF) from NODDI, and total apparent fiber density (AFDt) from CSD—as well as different brain regions, namely the middle section of the corpus callosum (CC\_Mid), the left corticospinal tract (CST\_L), and the right inferior fronto-occipital fasciculus (IFOF\_R).

\begin{figure}[h] % "t" option places the figure at the top of the page
    \centering
    \includegraphics[width=\linewidth]{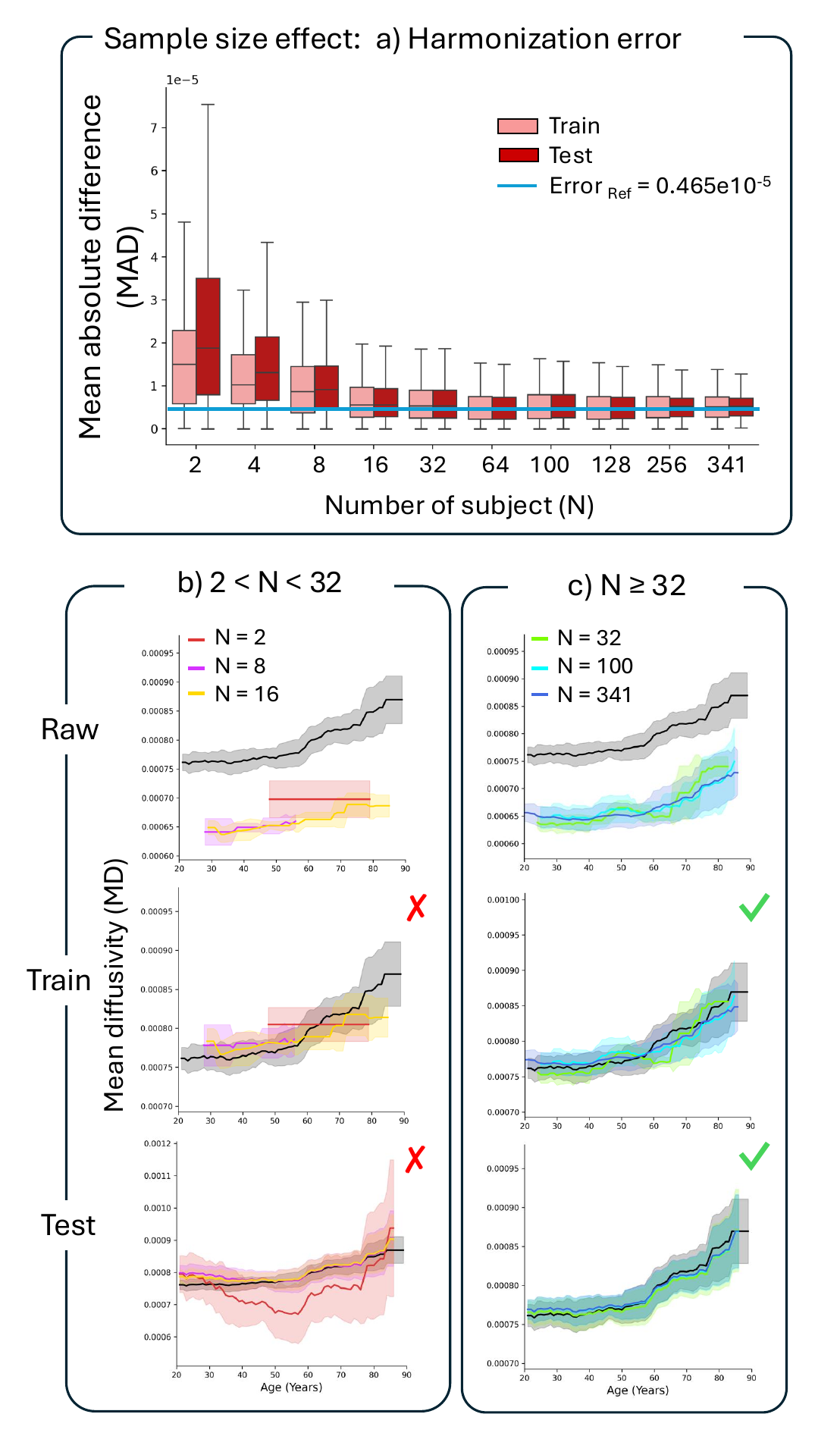}
    \caption{(a) Mean Absolute Difference (MAD) training and testing harmonization errors for various number of training samples $N$ from the Modified-CamCAN moving site. (b) and (c) illustrate the effect of using too few or enough training samples on the training and testing fit. Green checks indicate correct harmonization, red X's indicate poor harmonization.}\vspace{-0.4cm}
    \label{fig:Nsubjects_camcan}
\end{figure}

%%%%%%%%%%%%%%%%%%%%%%%%%%%%%%%%%%%%%%%%%%%%%%%%%%%%%%%%%%%%%%%%%%%%%%%%%%%%%%%%%%%%%%%%%%%%%%%%%%
\section{Results}
\label{sec:results}

\subsection{Experiment 1: Additive and multiplicative biases}

Figure~\ref{fig:factors_exp_camcan} illustrates the effect of the bias, slope, and variance multiplicative factors ($A$, $M$, and $S$) from Eq.~(\ref{eq:combat_M1_M2_A_Bias}) on ComBAT harmonization. As shown in column (a), ComBAT effectively compensates for bias effects when $M=S=1$. Across all cases, the harmonized population exhibits negligible Bhattacharyya distances (BD) on the order of $2.5 \times 10^{-5}$. However, as seen on column (b), ComBAT struggles to harmonize data from two sites with significantly different covariate slopes. While harmonization is adequate for a multiplicative slope factor close to 1 like $S=0.8$, the alignment between the red and black populations deteriorates dramatically for $S=0$ and $S=1.8$.

\begin{figure*}[tp] % "t" option places the figure at the top of the page
    \centering
    \includegraphics[width=\textwidth]{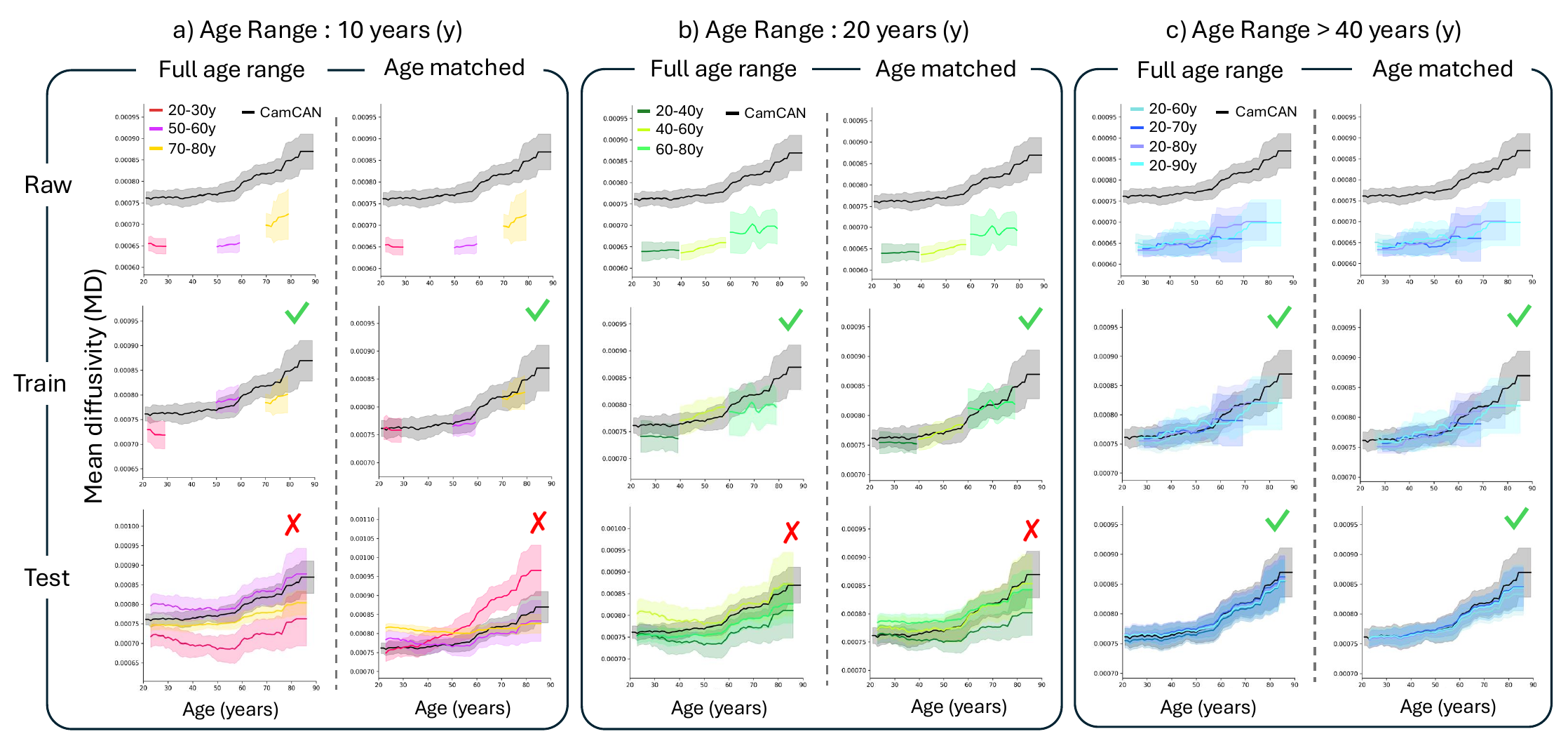}
    \caption{Effect of age range harmonization of the mean diffusivity (MD) metric between the modified CamCAN and CamCAN. Green checks indicate correct harmonization, red X's indicate poor harmonization.}
    \label{fig:agerange_camcan}
\end{figure*}

The third column (c) demonstrates that, in the absence of bias and slope factors ($A=S=1$), ComBAT performs well in compensating for low and high multiplicative variance factors (i.e., $M=0.2$ and $M=1.8$). However, as depicted in the bottom-right figures, when there is a difference in slope between the populations (i.e., $S=1.8$), ComBAT fails to accurately estimate the population variance $\hat{\delta}_{iv}^{2*}$.

To illustrate the impact of the slope multiplicative factor $S$ on ComBAT, Figure~\ref{fig:factors_exp_var-error} shows the quadratic error in estimating the moving site variance $\hat{\delta}_{iv}^{2*}$ for varying slope values $S \in [0, 2]$. Each curve represents a unique variance multiplicative factor $M$. Notably, ComBAT achieves zero error in variance estimation when $S=1$, i.e. when the covariate slopes are the same for both populations. However, as $S$ deviates from $1$ (either decreasing towards $0$ or increasing towards $2$), the quadratic error in $\hat{\delta}_{iv}^{2*}$ estimation rises.

A similar scenario is illustrated on the bottom of Figure~\ref{fig:factors_exp_var-error}, which shows the harmonization of the mean diffusivity (MD) of the NIMH population (red) onto that of CamCAN (gray). The difference in covariate slopes between NIMH and CamCAN (left) leads to erroneous harmonization outcomes (right).

\subsection{Experiment 2: Training sample size}

The results presented in Figure~\ref{fig:Nsubjects_camcan} explore the impact of the number of training subjects $N$ from the moving site on harmonization performance. Figure~\ref{fig:Nsubjects_camcan} (a) highlights the relationship between training and testing errors when harmonizing varying numbers of training subjects, ranging from 2 to 341, from the Modified-CamCAN dataset to the reference site. The dashed line in the figure represents the minimum achievable error, which is obtained when the entire dataset is utilized for harmonization.

When examining these results, several patterns become apparent. When a very small number of subjects ($N<8$) is used, the testing error is significantly higher than the training error. This discrepancy suggests overfitting, as the model struggles to generalize with limited data. As the sample size increases beyond $N=8$ subjects, the testing and training errors begin to decrease, indicating that ComBAT generalizes well even with moderate sample sizes. Importantly, although the training and testing errors approach the minimum achievable error when the sample size reaches $64$ subjects or more, it is evident that reasonable harmonization performance can already be achieved with a sample size between $16$ and $32$ subjects.

This trend is further illustrated in Figures~\ref{fig:Nsubjects_camcan}~(b) and (c), which provide additional insights into the relationship between sample size and harmonization quality. With $16$ or fewer subjects, the testing results remain inconsistent and relatively poor, underscoring the limitations of working with such small datasets. However, as the number of subjects increases to $32$ or more, both the training and testing results improve substantially, with harmonization achieving a strong fit to the data.

\subsection{Experiment 3: Training age range}
\begin{figure}[tp] % "t" option places the figure at the top of the page
    \centering
    \includegraphics[width=0.7\linewidth]{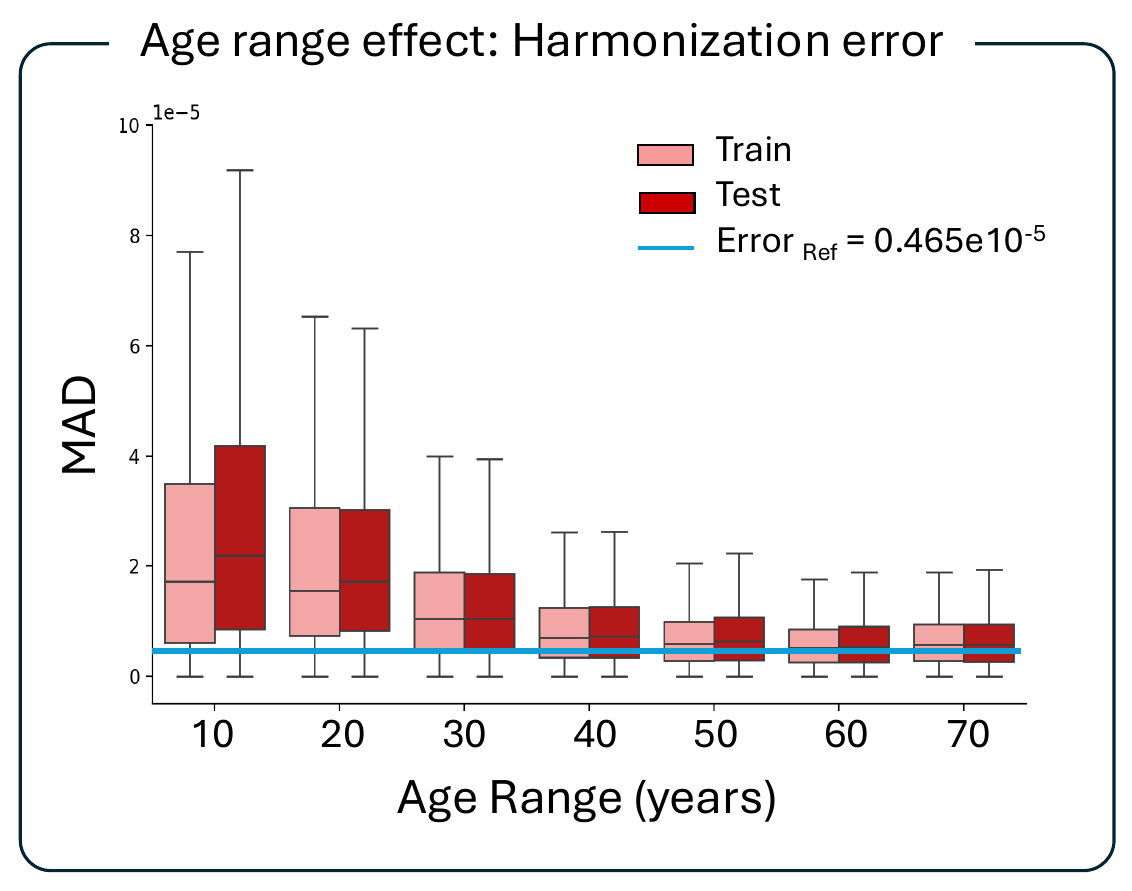}
    \caption{Training and testing mean absolute difference (MAD) harmonization error for age ranges from 10 to 70 years of the Modified-CamCAN dataset.  }
    \label{fig:Nsubjects_harmerror}
\end{figure}

The harmonization of the Modified-CamCAN moving dataset across different age ranges is depicted in Figure~\ref{fig:Nsubjects_harmerror} and ~\ref{fig:agerange_camcan}.  

Figure~\ref{fig:Nsubjects_harmerror} reports the training and testing MAD for age ranges from $10$ to $70$ years. These results emphasize that as the age range widens, the harmonization performance improves, with both training and testing errors stabilizing at a low plateau for age ranges of $40+$ years. This suggests that broader age ranges provide more robust estimates of harmonization parameters, improving generalization for unseen data.

More specific harmonization examples for different age ranges are depicted in Figure~\ref{fig:agerange_camcan}. From left to right, the panels present results for age ranges spanning $10$, $20$, and $40$ years. The first row illustrates unharmonized data for three specific age groups ($10-20$, $50-60$, and $70-80$ years in Figure~\ref{fig:agerange_camcan}(a)), while the second and third rows show the harmonized training and testing data, respectively. In all cases, harmonization outcomes are compared between two approaches: using the full CamCAN age range ($20-90$ years) or restricting to CamCAN subjects whose ages match those of the moving site (i.e., $10-20$, $50-60$, and $70-80$ years for Figure~\ref{fig:agerange_camcan}(a)).

Across all settings, ComBAT achieves relatively low training errors, particularly when harmonization is performed using reference site subjects whose ages align closely with those in the moving site. This effect is especially pronounced for the $20-30$ and $70-80$ year age ranges, where the harmonized training distribution deviates less from the original black CamCAN distribution. However, examining the testing results reveals a critical limitation: smaller age ranges, such as $10$ or $20$ years, often result in poor harmonization fits for the testing data. Despite this, the results in Figure~\ref{fig:agerange_camcan}(c) demonstrate that good training and testing harmonization fits can be achieved when the age range exceeds $40$ years.

\begin{figure}[!t]
    \centering
    \includegraphics[width=0.99\linewidth]{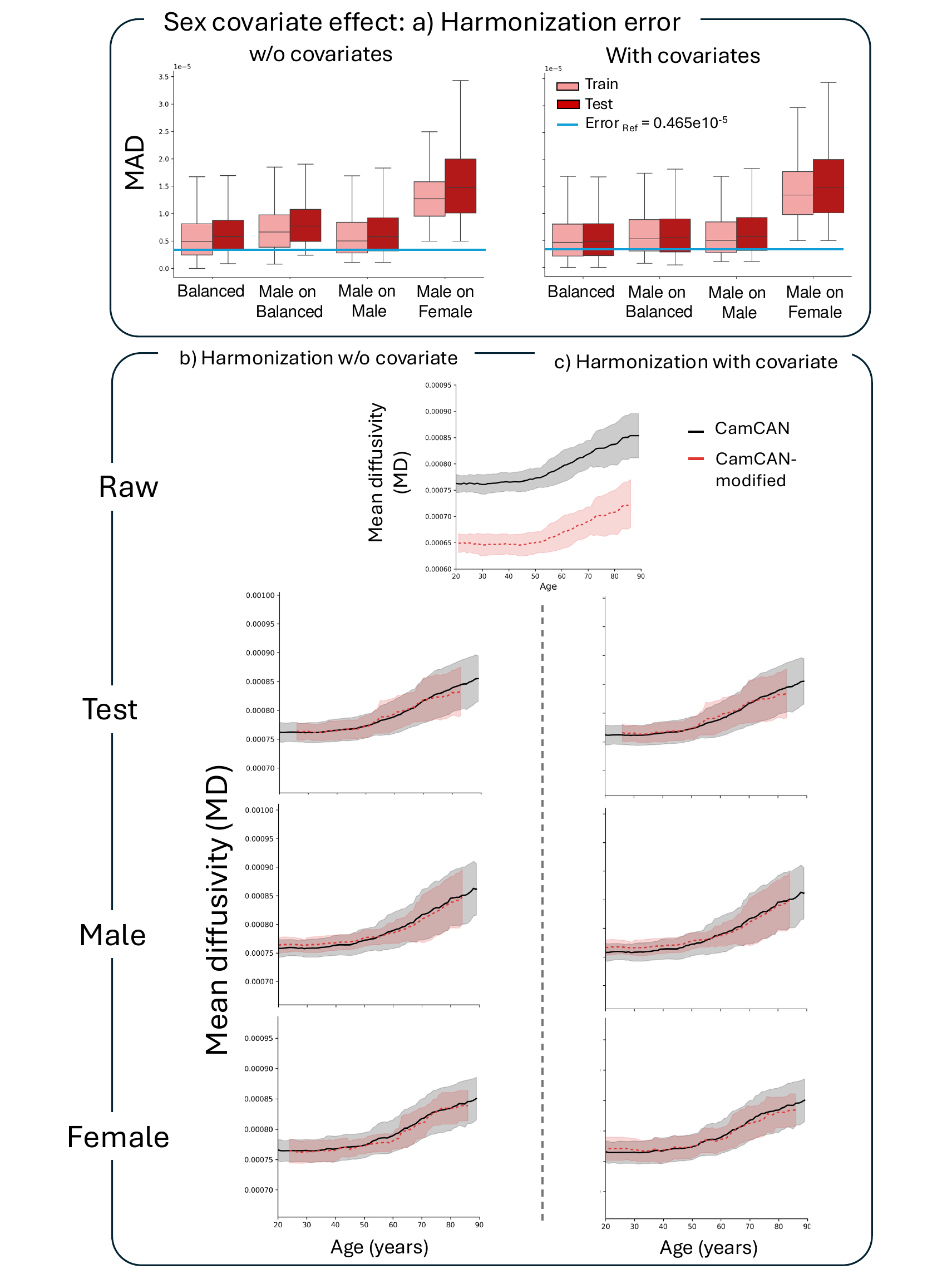}
    \caption{Effect of covariate on Pairwise-ComBAT harmonization of the mean diffusivity (MD) metric between modified CamCAN and CamCAN datasets. a) Mean Absolution Difference (MAD) harmonization when considering male-only, female-only and a balance (male+female) populations. b) and c) harmonization results when using two balanced populations without (b) and with (c) sex covariate.}
    \label{fig:gendercov_camcan}
\end{figure}

\begin{figure*}[tp]
    \centering
    \includegraphics[width=\textwidth]{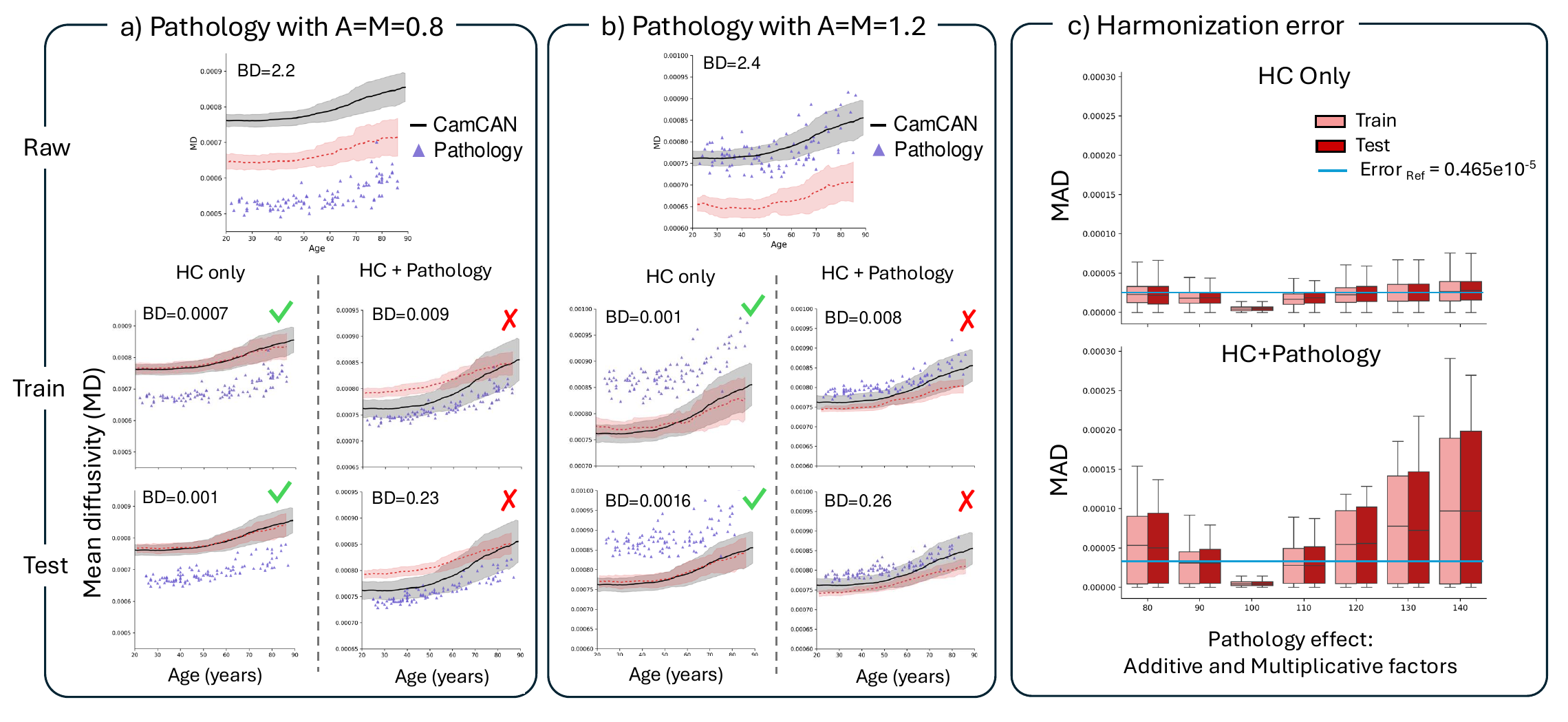} \vspace{-0.4cm}
    \caption{(a) and (b) illustrate a pathological population (purple triangles) exhibiting negative (top left) and positive (middle top) bias relative to a normative population (red). Below these, the harmonization results are shown both with and without the inclusion of pathological cases in the training dataset. Green checks indicate correct harmonization, red X's indicate poor harmonization. (c) presents the harmonization mean absolution difference (MAD) error, comparing scenarios with [bottom] and without [top] the inclusion of pathological cases during training. }
    \label{fig:patho_camcan}
\end{figure*}

\begin{figure}[!t]
    \centering
    \includegraphics[width=\linewidth]{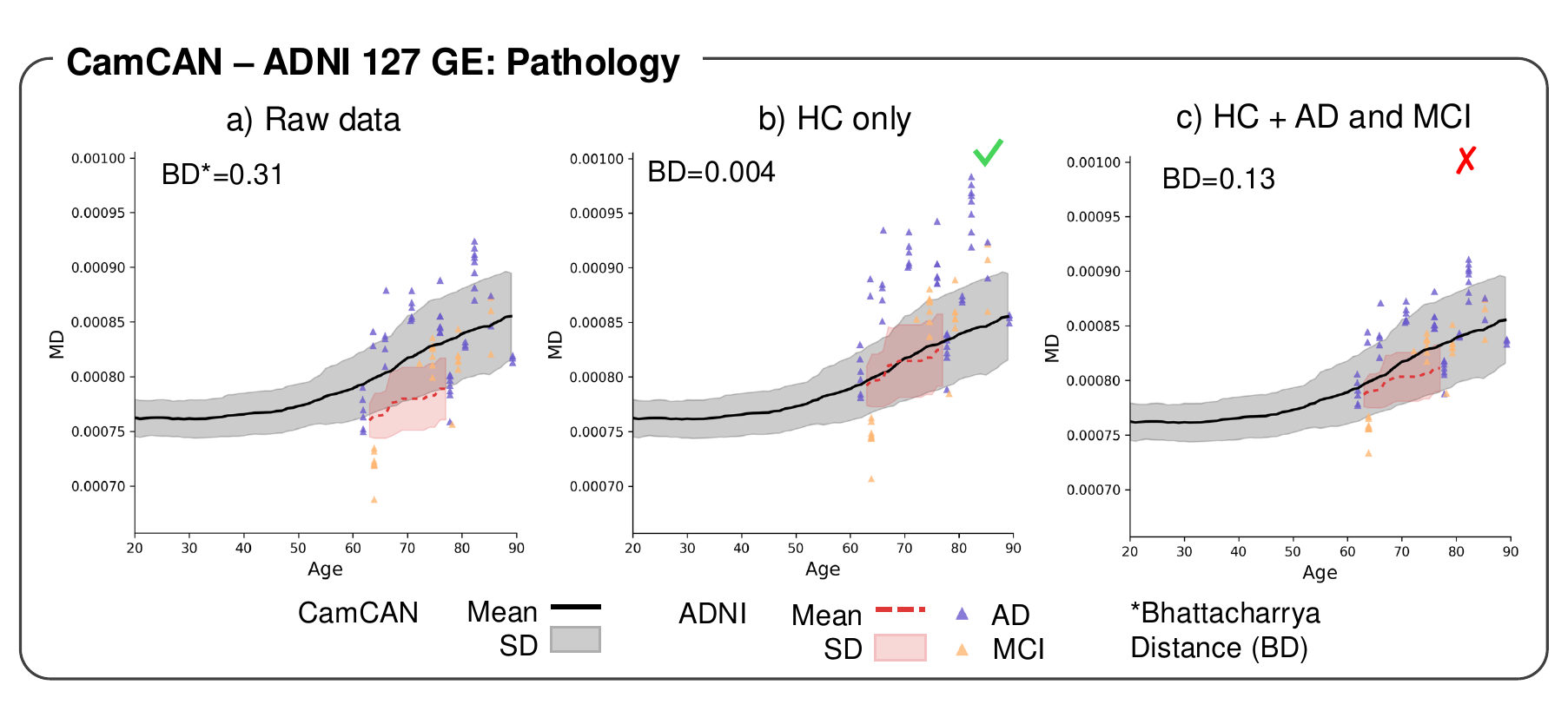}
    \caption{Harmonization of ADNI-127-GE site with CamCAN. a) Raw data, b) Pairwise-ComBAT harmonization from HC group only, c) Pairwise-ComBAT harmonization from HC and pathological group. The gray (target site) and red (moving site) lines show the population mean trend of both sites, and the corresponding light color indicates the standard deviation. Green checks indicate correct harmonization, red X's indicate poor harmonization.}
    \label{fig:patho_adni}
\end{figure}

\subsection{Experiment 4: Sex covariate}
\label{subsec:res_covgender}

The top panel of Figure~\ref{fig:gendercov_camcan} illustrates the MAD harmonization error across various scenarios: harmonizing two balanced populations (i.e., equal numbers of males and females), harmonizing a male-only population with a balanced population, harmonizing two male-only populations, and harmonizing a male-only population with a female-only population. Additionally, we assessed the effect of including the sex covariates ($x_{ij}$ and $\beta_v$) in the ComBAT equations. The left side of the figure shows results where the sex covariates were excluded from the harmonization equations, while the right side shows results with these covariates included.

Interestingly, the first three scenarios result in low training and testing harmonization errors, while the fourth—harmonizing a male-only population with a female-only population—exhibits significantly larger errors. While accounting for sex covariates slightly improves the harmonization of a male-only population with a balanced population, this improvement is subtle. These findings emphasize the challenges and limitations of harmonizing highly heterogeneous populations, where covariate adjustments alone may not suffice.

The bottom panel of Figure~\ref{fig:gendercov_camcan} shows the results of harmonizing two balanced populations with and without incorporating sex covariates in the harmonization equations. The progression begins with the raw data, followed by the harmonized balanced test set, and concludes with visualizations of the harmonized male and female populations. Notably, including the sex covariate in the ComBAT equation does not significantly impact the results when the populations are well-balanced in terms of male and female representation.

\subsection{Experiment 5: Pathological populations}
\label{subsec:res_patho}

Figures~\ref{fig:patho_camcan} and \ref{fig:patho_adni} present harmonization results when the moving site includes both healthy control (HC) subjects and subjects affected by a pathology. In Figure~\ref{fig:patho_camcan} (a) and (b), the pathological cases (blue dots) are a subset of subjects from the Modified-CamCAN dataset, where a negative bias is introduced in (a) and a positive bias in (b). In both scenarios, the harmonized training and testing data highlight the critical importance of using only HC subjects for harmonization onto a reference, such as CamCAN. Including pathological cases in the harmonization process results in a compressed population, where the harmonized HC and pathological cases are erroneously aligned with the target population (gray).

Additionally, Figure~\ref{fig:patho_camcan} (c) demonstrates that including pathological cases in the training set significantly increases the harmonization error. This issue becomes particularly problematic when the pathology's bias falls below $A=M=0.9$ or exceeds $A=M=1.1$.   

Figure~\ref{fig:patho_adni} illustrates the HC subjects (red-shaded region) and pathological cases (AD and MCI markers) from the 127-GE ADNI site in comparison with the CamCAN dataset. When the harmonization parameters are computed using only HC subjects, the red and gray-shaded populations are perfectly aligned, and the pathological cases remain appropriately distinct as outliers. Conversely, if both HC and pathological cases are used to estimate the ComBAT harmonization parameters, the pathological population is compressed into the normative range, creating the false impression that these subjects are normal.

\vspace{-0.3cm}
\section{Discussion}
\label{sec:discussion}

We evaluated the often-overlooked influence of various factors on the effectiveness of the ComBAT harmonization method, using a pairwise implementation of the technique. Our experiments investigated the effects of site-wise biases, sample size, age range, the sex covariate, and the presence of pathologies.  Results demonstrate that ComBAT’s performance can be impacted by these factors, which underlines how ComBAT cannot be used as a black box. 

As a general guideline, ComBAT performs well when harmonizing relatively large, well-balanced populations (in terms of age and sex) with substantial inter-population overlap in age distributions and similar age covariate slopes.  ComBAT performs also well for compensating additive and multiplicative biases (parameters $A$ and $M$ through our experiments). Our findings also reiterate ComBAT’s ability to preserve the biological variability of critical covariates, such as age, while correcting for inter-site differences.

However, ComBAT cannot be used indiscriminately, as it may produce suboptimal harmonization under certain conditions. One prominent issue arises when the age covariate slope differs between the populations being harmonized. In such cases, while the additive bias is compensated, the multiplicative bias is incorrectly estimated, and the slope discrepancy remains uncorrected. As discussed in \S~\nameref{sec:limitations}, this limitation stems from the assumption that the slope $\beta_v$ is consistent across all sites, which is not always valid \cite{Bayer2022, Kim2024}.

Another challenge occurs when the sample size is too small for the moving site being harmonized to the reference site. Fortin et al. \cite{Fortin2017}, the first to apply ComBAT to dMRI data harmonization, recommended a minimum of $N=20$ subjects per site for reliable harmonization. Similarly, Orlhac et al. \cite{Orlhac2022} reported that ComBAT becomes less reliable for harmonization of PET imaging biomarkers across two sites when sample sizes fall below $N=30$ subjects per site. Our results are consistent with these studies, showing that a reasonable harmonization can already be obtained with $16$ to $32$ subjects and that fewer subjects many result into overfitting (low training error and large testing error).

Another issue emerges when the age range of a population is too narrow ($<20$ years). When reference subjects closely match the age range of the moving site (i.e., $10$-$20$ or $70$-$80$), training errors remain relatively low, indicating a good fit to the training data. However, this age-matched approach of adjusting reference distributions based on moving site characteristics does not necessarily result in optimal generalization, as shown by high test errors. This is particularly the case when the age range is restricted to $10$ or $20$ years, for which test performance deteriorates, leading to inconsistencies in the harmonization, especially for age ranges not included in the training data. In contrast, the improvement in both training and test errors is reached with age ranges of $40$ years or greater.  
This finding is consistent with previous research suggesting that wider age distributions and age range overlap are required for an optimal ComBAT harmonization \cite{Pomponio2020, Beer2020}.

Regarding the covariates, Pairwise-ComBAT assumes that all subjects share the same distribution when covariates are not included, i.e. it assumes identical distributions between males and females, which risks suppressing biologically relevant variability by wrongly attributing it to site effects \cite{Kim2024, Orlhac2022}. The inclusion of covariates helps to account for differences in distributions, thereby improving harmonization - especially in cases of demographic imbalance. This is consistent with the findings of Kim et al. \cite{Kim2024}, who reported that ComBat struggles to harmonize sites with very small and demographically diverse samples unless appropriate covariates are included.

Results also emphasize the importance of handling pathological populations carefully. In clinical studies of brain diseases like Alzheimer's (AD), Multiple Sclerosis (MS) or brain tumours, the site effect is often influenced by differences in clinical status \cite{Kim2024, Orlhac2022, shin2017}. To prevent site effects from confounding biological variability, previous studies have typically included clinical status as a covariate \cite{Kim2024, shin2017}. In contrast, this study excludes clinical status from the estimation of training parameters, rather than using it as a covariate. When pathological and healthy subjects are mixed, Pairwise-ComBAT tends to compress both pathological and healthy subjects around the target site, thereby reducing population-related variability. The proposed harmonization strategy preserves biological variability while correcting for site effects, thus avoiding bias in the pathological population. Thereby, a recommended strategy is to estimate the harmonization parameters using only control subjects, temporarily excluding pathological cases. Once estimated, there parameters can then be applied to the pathological subjects. This approach minimizes the risk of confounding disease-related variability with site effects \cite{Orlhac2022}.

In conclusion, our findings underscore that the key to successful ComBAT harmonization lies in ensuring that populations are as homogeneous as possible in terms of age range, demographic profiles, sex distribution, covariate slope and health status. Neglecting these considerations may not only lead to poor harmonization during training but also result in catastrophic testing errors.

\subsection{Recommendations for a good ComBAT harmonization}
\label{subsec:recommendations}

Given the influence of the factors discussed above, we strongly recommend carefully examining the characteristics of the populations from each site before proceeding with their harmonization. This step ensures that ComBAT's assumptions are met, particularly concerning observed distributions (i.e., additive and multiplicative effects) and sample size, and also helps identify one or more relevant covariates. These covariates can be demographic (i.e., age, sex, or handedness) or clinical (i.e., neurodegenerative disease, treatment, or the presence of a brain abnormality, such as a tumor). 

In the light of our experiments, our recommendations are as follows:

\begin{itemize}
    \item \textbf{Data inspection:}  
    ComBAT should never be used as a "magic black box." Always ensure that the populations to be harmonized exhibit distributions that align with ComBAT’s assumptions, particularly uniformity across key variables. \vspace{-0.0cm}
    \item \textbf{Covariate slope effect:}  
    The data slope $\beta_v$ of key co-variates such as age should be approximately the same across all sites. If this condition is not met, the resulting harmonization is likely to be erroneous. \vspace{-0.0cm}
    \item \textbf{Sample size:}  
    At least 16 to 32 subjects should be included when computing the parameters of ComBAT to ensure that the resulting harmonization function generalizes well to all data.\vspace{-0.0cm}
    \item \textbf{Age distributions:}  
    The age range of the moving site's population should span at least 40 years. If this requirement cannot be met, consider using an age-matched ComBAT harmonization, particularly for applications where proper harmonization of the training population is crucial.\vspace{-0.0cm}

    \item \textbf{Sex covariate:}  
    Populations from all sites should have a uniform distribution of male and female subjects. Under no circumstances should a male predominant population be harmonized with a female predominant population, and vice versa.\vspace{-0.0cm}

    \item \textbf{Disease differences:}  
    For normative modeling applications, always compute the ComBAT harmonization parameters using a group of healthy subjects. These parameters can later on be applied to pathological subjects, reducing the risk of confounding disease-related variability with site effects.
\end{itemize}

Interestingly, some of our recommendations echos  that of Karayumak et al \cite{Cetin-Karayumak2019} (LinearRISH method) which mention that : "{\em At least 16 to 18 well-matched (sex and age) healthy controls from each site are needed to reliably capture site related differences}".

\vspace{-0.2cm}
\section{Conclusions and future works}
\label{sec:conclusions}
This study rigorously examined the ComBAT harmonization methodology and its variant, Pairwise-ComBAT, in addressing biases across multi-site diffusion MRI datasets. The findings highlight the critical importance of adhering to foundational assumptions, such as demographic consistency, sufficient sample size, and proper covariate representation, to ensure robust harmonization outcomes. While ComBAT demonstrated remarkable efficacy in compensating for additive and multiplicative biases, challenges arise when site-wise demographic distributions or covariate slopes differ significantly. We also provided six recommendations for an optimal application of ComBAT in both research and clinical settings, emphasizing its potential for preserving biological variability and facilitating reproducible science.

\vspace{-0.3cm}
\section*{Code and Data Availability}
\vspace{-0.2cm}
The source code and documentation of the Pairwise-ComBAT method as well as the data used in the experimental section will be rendered public upon acceptance of this paper. %is available at \url{https://github.com/scil-vital/Vanilla-ComBAT}.
%All MRI dataset used in this work are publically available.

%\section*{Funding}
%This work was funded by ...
\vspace{-0.3cm}
\section*{Acknowledgments}
This work was supported by the ACUITY/CQDM consortium, the Natural Sciences and Engineering Research Council of Canada throught its discovery grant program, and the MITACS Acceleration funding program. We are also thankful for the Institutional Université de Sherbrooke Research Chair in Neuroinformatics and NSERC Discovery Grant. 

Data collection and sharing for this project was funded by the Alzheimer's Disease Neuroimaging Initiative (ADNI) (National Institutes of Health Grant U01 AG024904) and DOD ADNI (Department of Defense award number W81XWH-12-2-0012). ADNI is funded by the National Institute on Aging, the National Institute of Biomedical Imaging and Bioengineering, and through generous contributions from the following: AbbVie, Alzheimer’s Association; Alzheimer’s Drug Discovery Foundation; Araclon Biotech; BioClinica, Inc.; Biogen; Bristol-Myers Squibb Company; CereSpir, Inc.; Cogstate; Eisai Inc.; Elan Pharmaceuticals, Inc.; Eli Lilly and Company; EuroImmun; F. Hoffmann-La Roche Ltd and its affiliated company Genentech, Inc.; Fujirebio; GE Healthcare; IXICO Ltd.; Janssen Alzheimer Immunotherapy Research \& Development, LLC.; Johnson \& Johnson Pharmaceutical Research \& Development LLC.; Lumosity; Lundbeck; Merck \& Co., Inc.; Meso
Scale Diagnostics, LLC.; NeuroRx Research; Neurotrack Technologies; Novartis Pharmaceuticals
Corporation; Pfizer Inc.; Piramal Imaging; Servier; Takeda Pharmaceutical Company; and Transition
Therapeutics. The Canadian Institutes of Health Research is providing funds to support ADNI clinical sites
in Canada. Private sector contributions are facilitated by the Foundation for the National Institutes of Health
(www.fnih.org). The grantee organization is the Northern California Institute for Research and Education,
and the study is coordinated by the Alzheimer’s Therapeutic Research Institute at the University of Southern
California. ADNI data are disseminated by the Laboratory for Neuro Imaging at the University of Southern
California

% \bibliographystyle{plain}
%\bibliography{reference}
\footnotesize

\printbibliography

\newpage
.
\newpage
\onecolumn
\section{Supplementary material}
This document contains complementary results to those presented in the paper.

\begin{figure*}[h]
    \centering
    \includegraphics[width=0.99\textwidth]{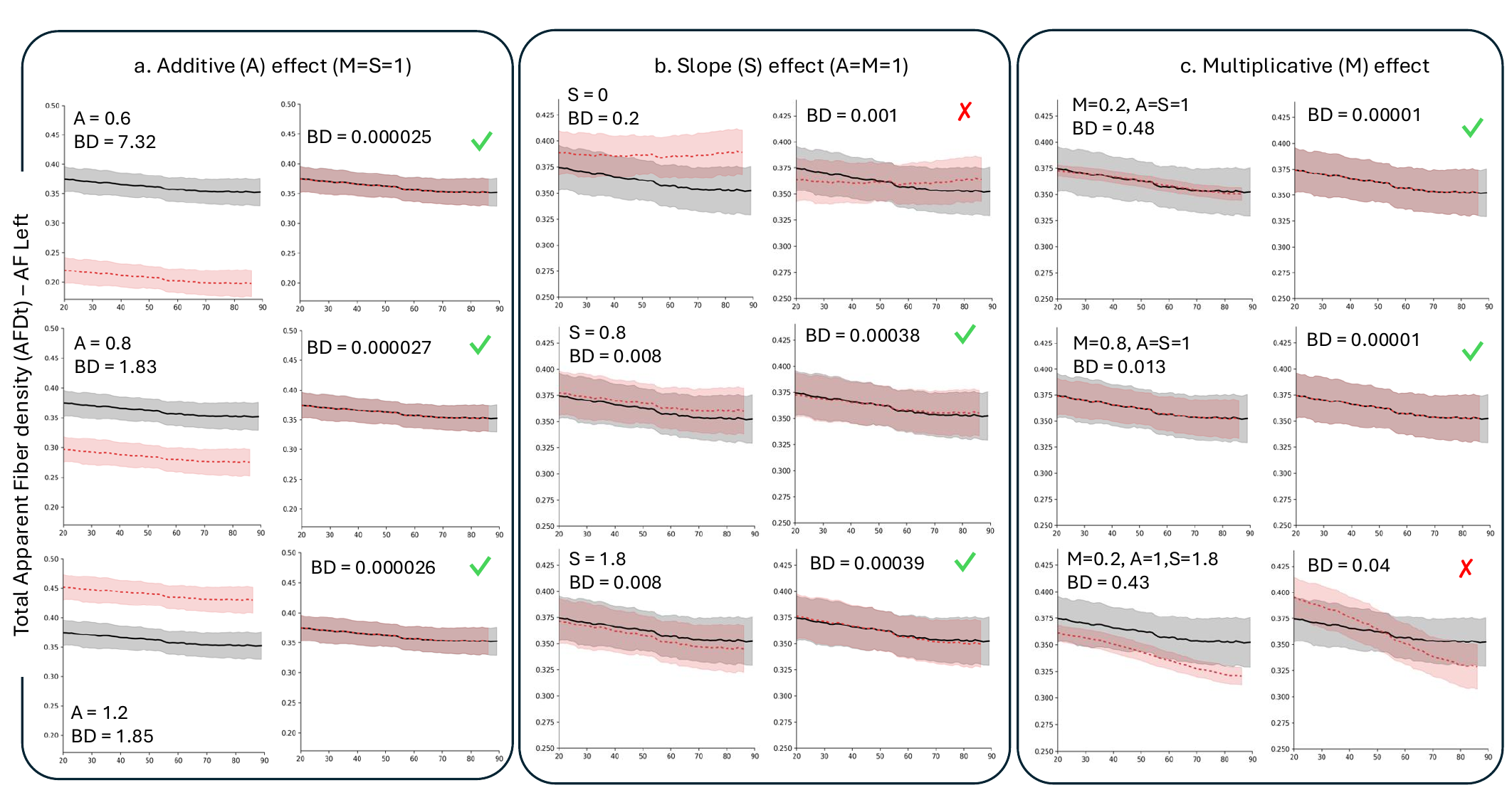} 
    \caption{This illustration is complementary to figure 3 in the paper.}
    \label{fig:S1}
\end{figure*}

\begin{figure}[h]
    \centering
    \includegraphics[width=0.99\textwidth]{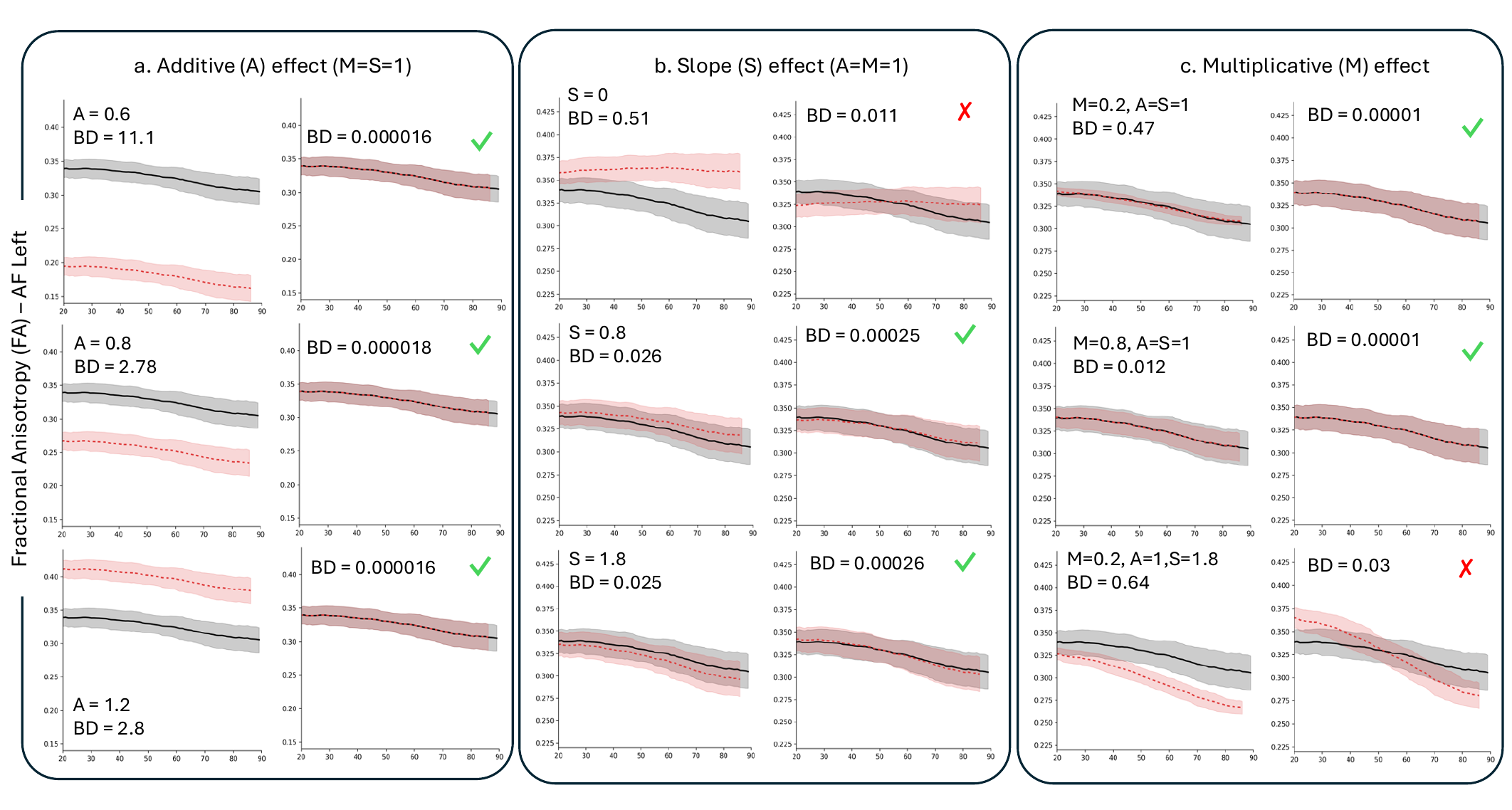} 
    \caption{This illustration is complementary to figure 3 in the paper.}
    \label{fig:S2}
\end{figure}

\begin{figure}[h]
    \centering
    \includegraphics[width=0.99\textwidth]{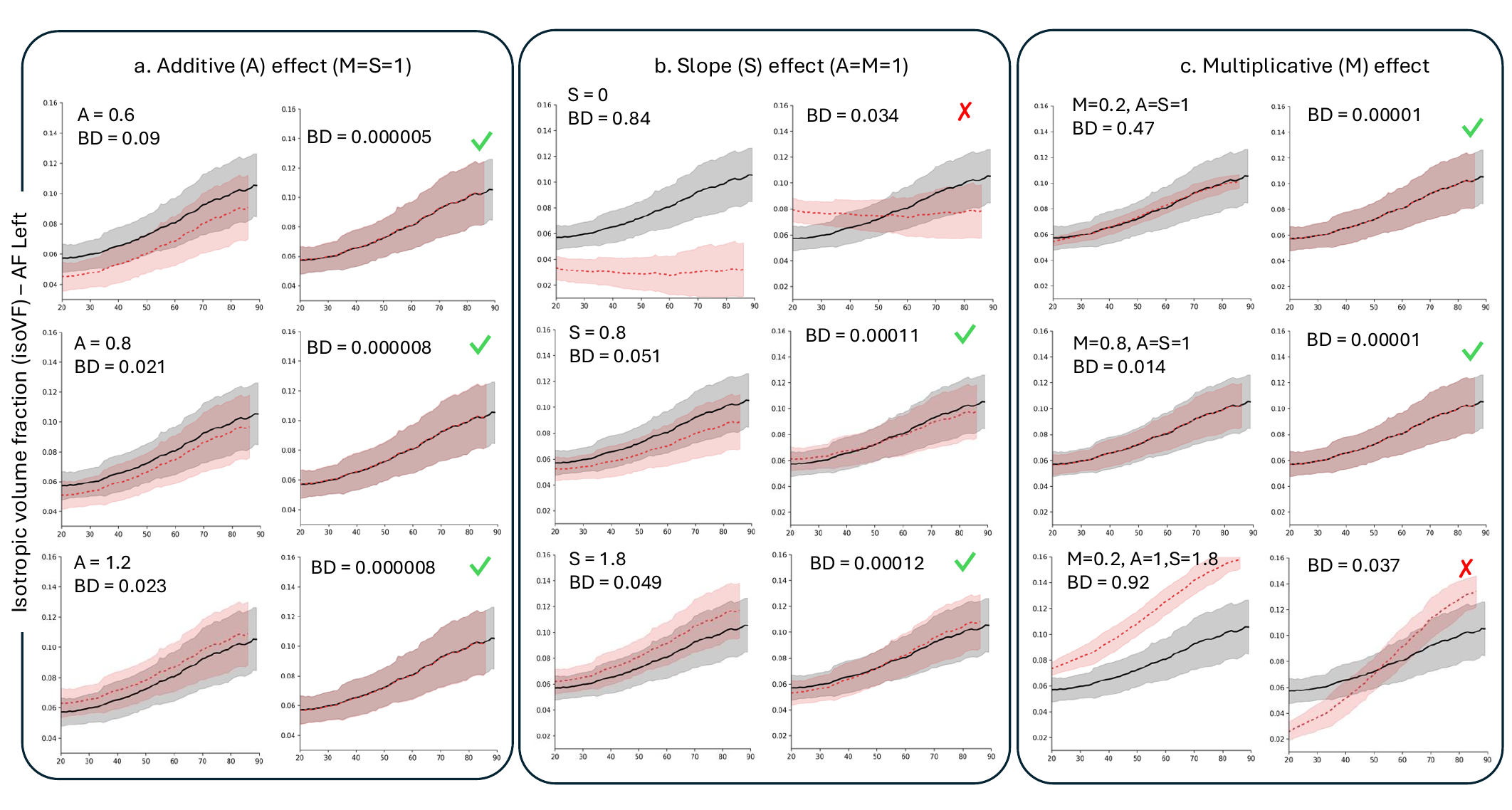} 
    \caption{This illustration is complementary to figure 3 in the paper.}
    \label{fig:S3}
\end{figure}

\begin{figure}[h]
    \centering
    \includegraphics[width=0.99\textwidth]{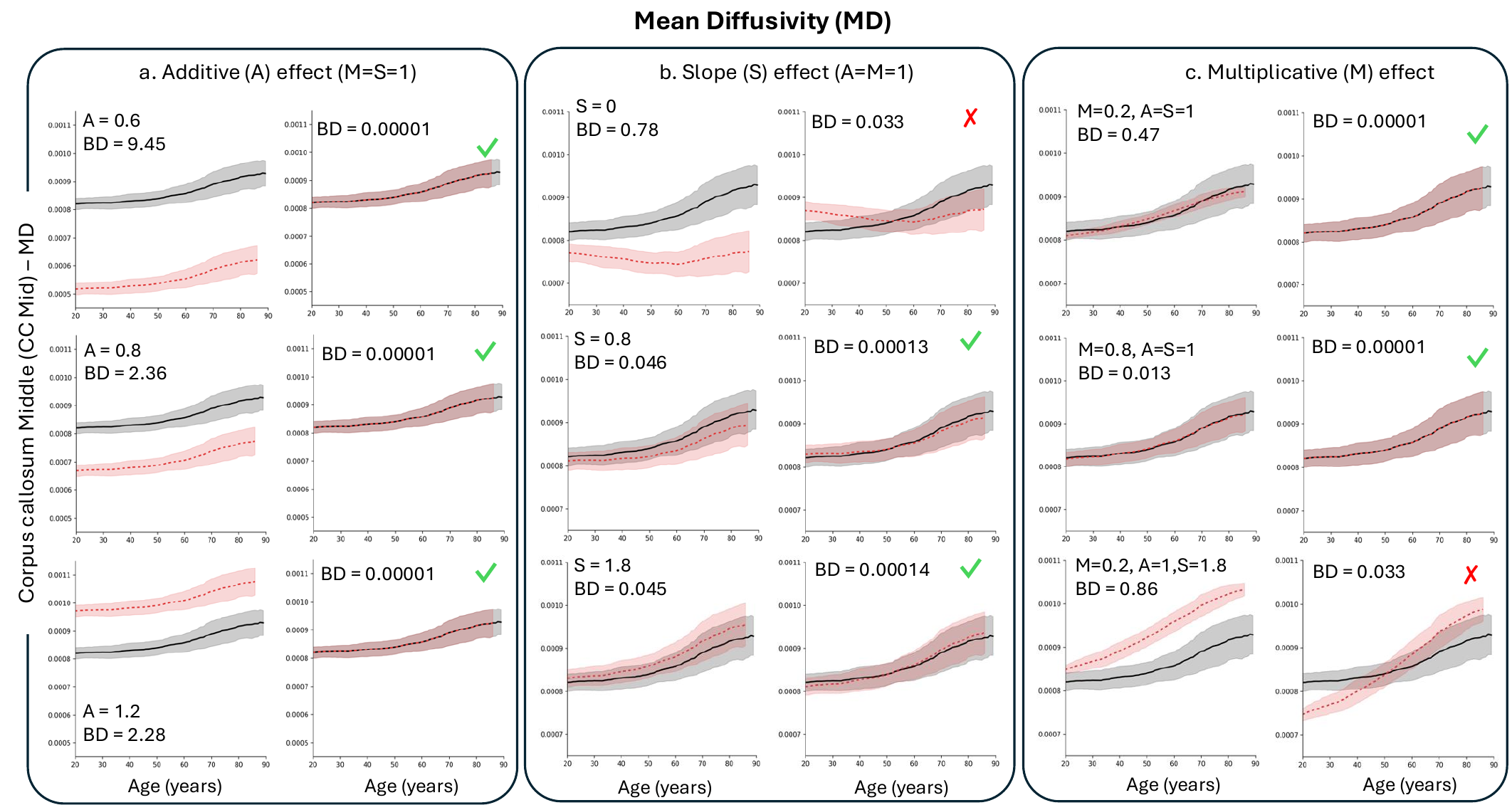} 
    \caption{This illustration is complementary to figure 3 in the paper.}
    \label{fig:S4}
\end{figure}

\begin{figure}[h]
    \centering
    \includegraphics[width=0.99\textwidth]{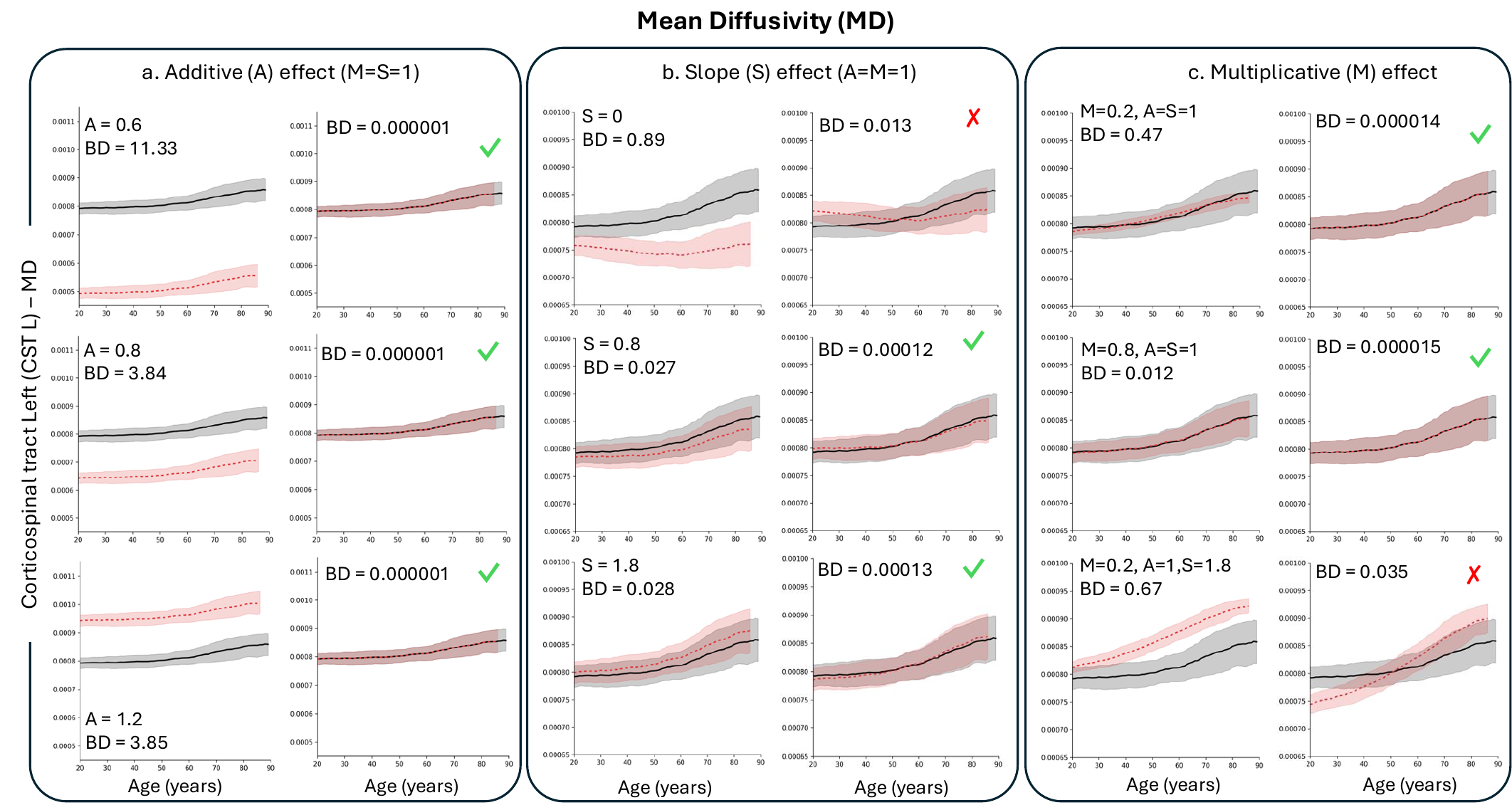} 
    \caption{This illustration is complementary to figure 3 in the paper.}
    \label{fig:S5}
\end{figure}

\begin{figure}[h]
    \centering
    \includegraphics[width=0.99\textwidth]{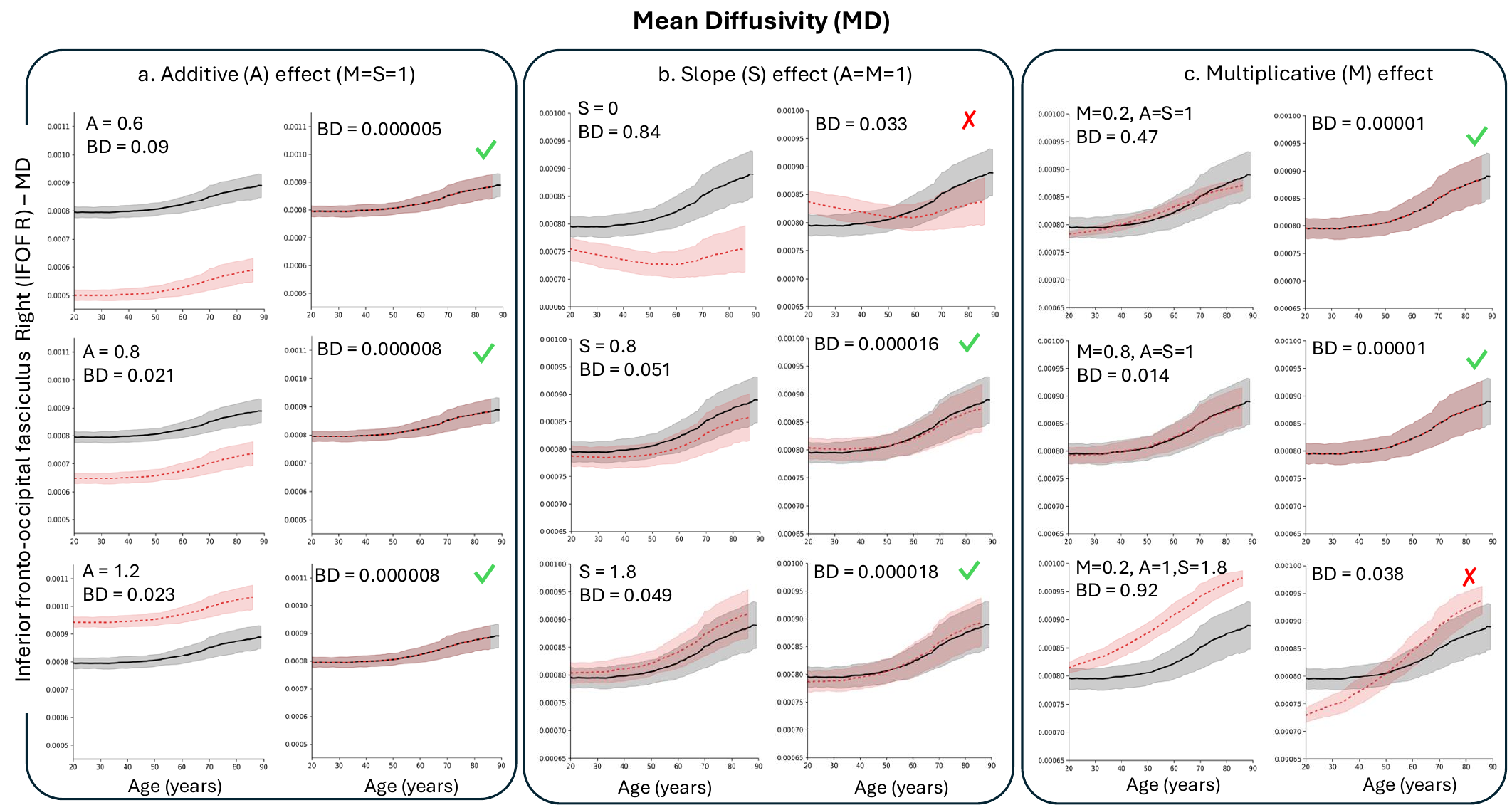} 
    \caption{This illustration is complementary to figure 3 in the paper.}
    \label{fig:S6}
\end{figure}

\paragraph{Figures~\ref{fig:S1} to \ref{fig:S6} complement Figure 3 in the paper.} Figures \ref{fig:S1} to \ref{fig:S3} illustrate the effect of variations in the slope of the covariate on three metrics—fractional anisotropy (FA), isotropic volume fraction (isoVF), and apparent fiber density (AFD)—measured in the left arcuate fasciculus (AFL). Figures \ref{fig:S4} to \ref{fig:S6} depict the same experiment but for mean diffusivity (MD), as in Figure 3, measured in the middle corpus callosum (CC\_Mid), the left corticospinal tract (CST\_L), and the right inferior fronto-occipital fasciculus (IFOF\_R).

\begin{figure}[h]
    \centering
    \includegraphics[width=0.99\textwidth]{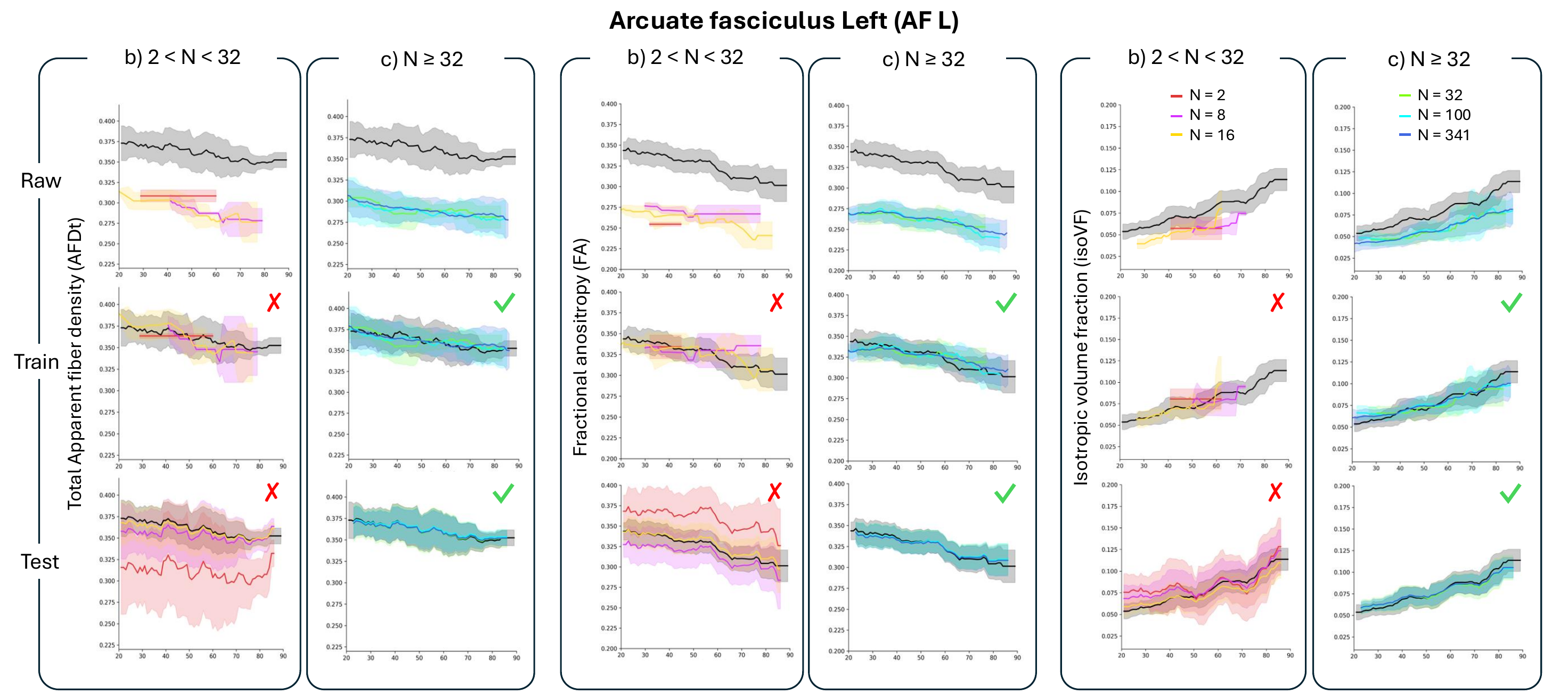} 
    \caption{This illustration is complementary to figure 5 in the paper.}
    \label{fig:S7}
\end{figure}

\begin{figure}[h]
    \centering
    \includegraphics[width=0.99\textwidth]{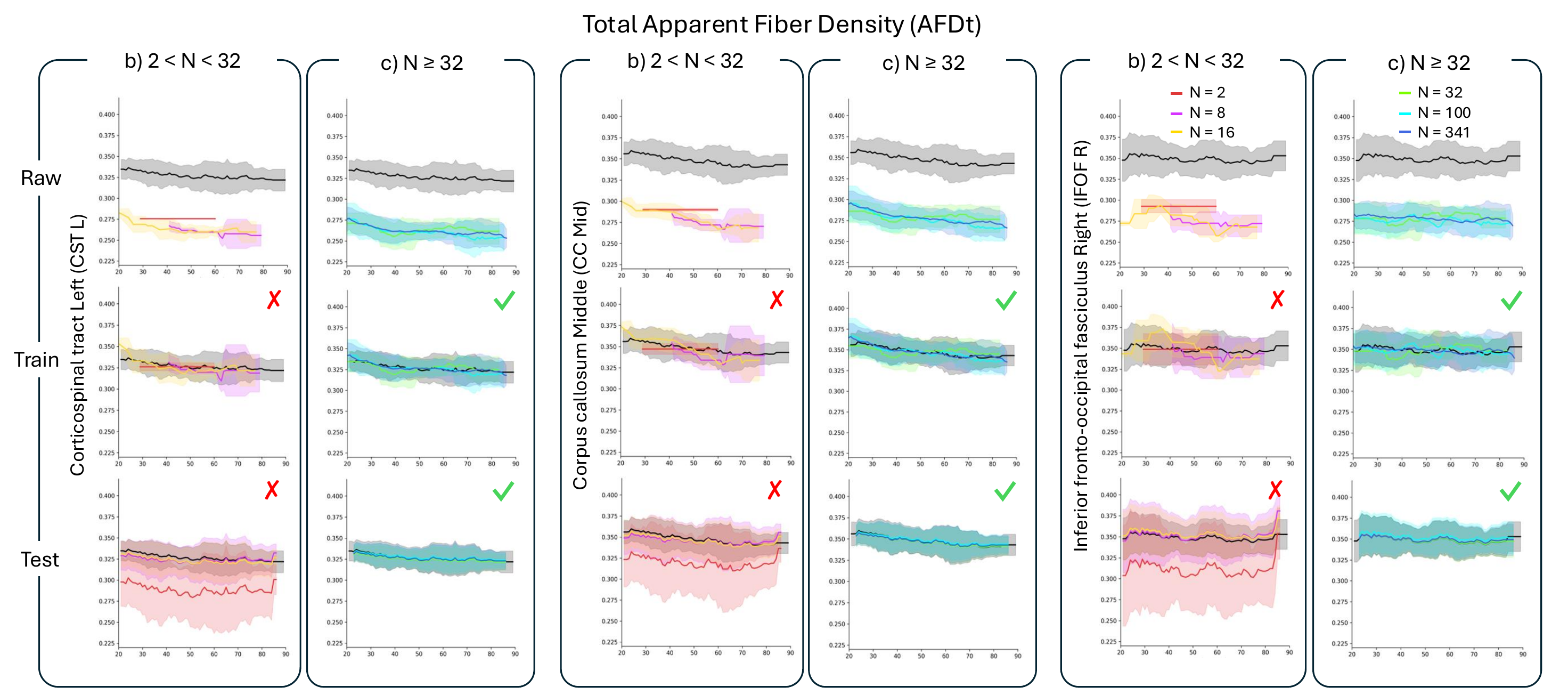} 
    \caption{This illustration is complementary to figure 5 in the paper.}
    \label{fig:S8}
\end{figure}

\paragraph{Figures \ref{fig:S7} and \ref{fig:S8} relate to Figures 5b and 5c in the paper.} Figure \ref{fig:S7} shows the effect of using a small number of samples (fewer than 32) versus a larger number of samples (32 and above) on the AFD, FA, and isoVF metrics measured in the AFL white matter region. Figure \ref{fig:S8} presents similar results for the AFD metric measured in the CST\_L, CC\_Mid, and IFOF\_R white matter regions.

\begin{figure}[h]
    \centering
    \includegraphics[width=0.99\textwidth]{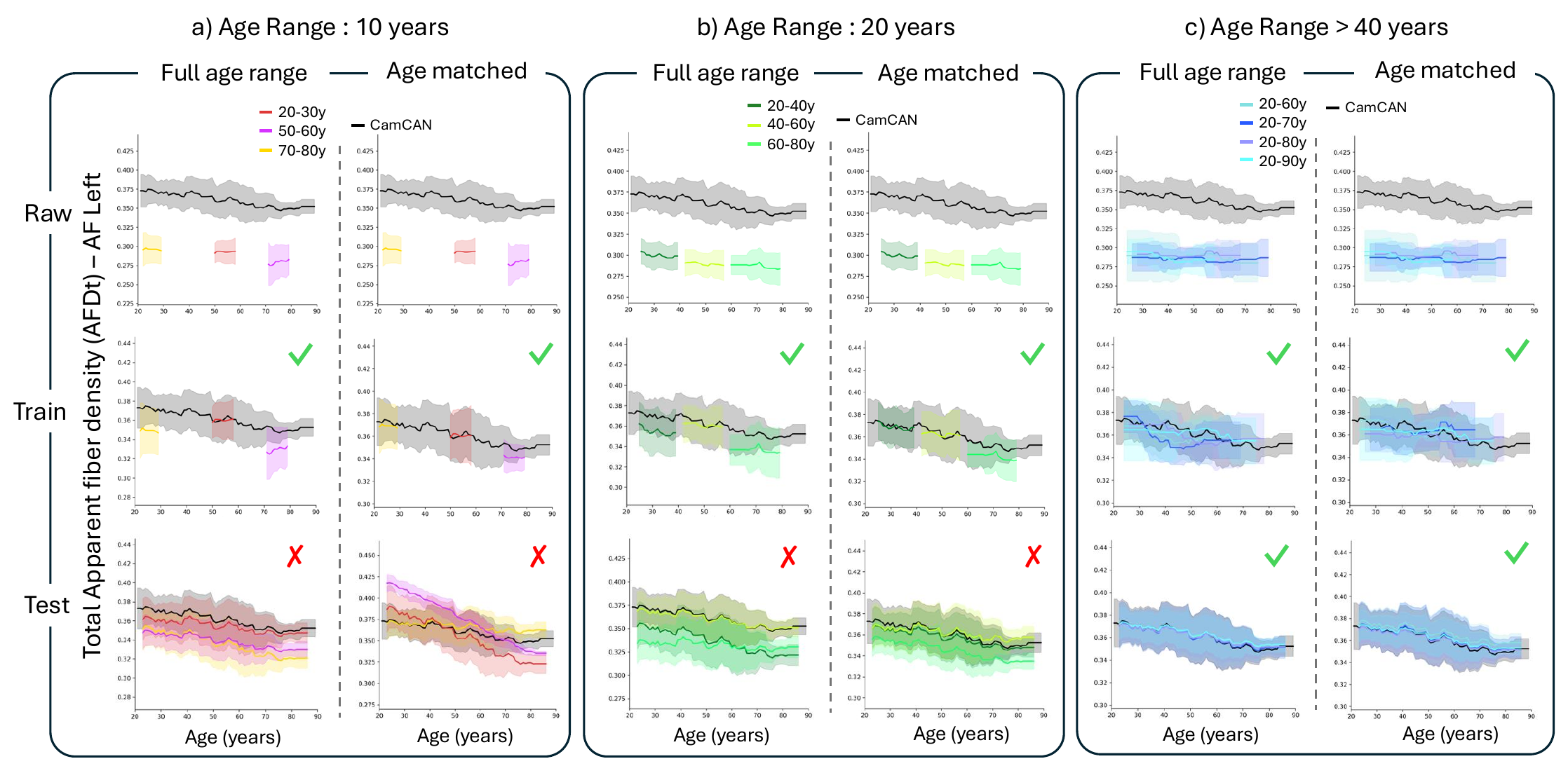} 
    \caption{This illustration is complementary to figure 6 in the paper.}
    \label{fig:S9}
\end{figure}

\begin{figure}[h]
    \centering
    \includegraphics[width=0.99\textwidth]{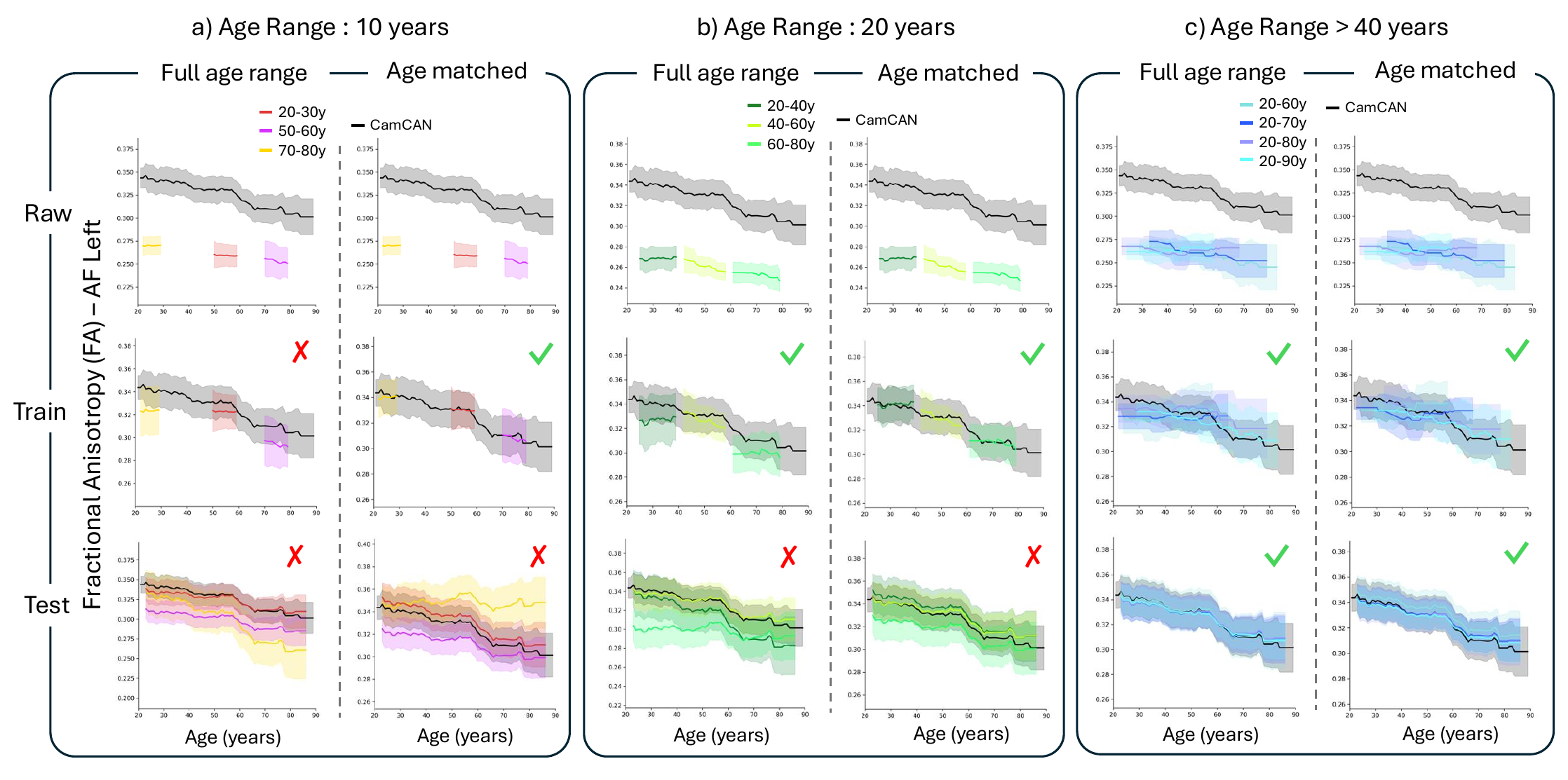} 
    \caption{This illustration is complementary to figure 6 in the paper.}
    \label{fig:S10}
\end{figure}

\begin{figure}[h]
    \centering
    \includegraphics[width=0.99\textwidth]{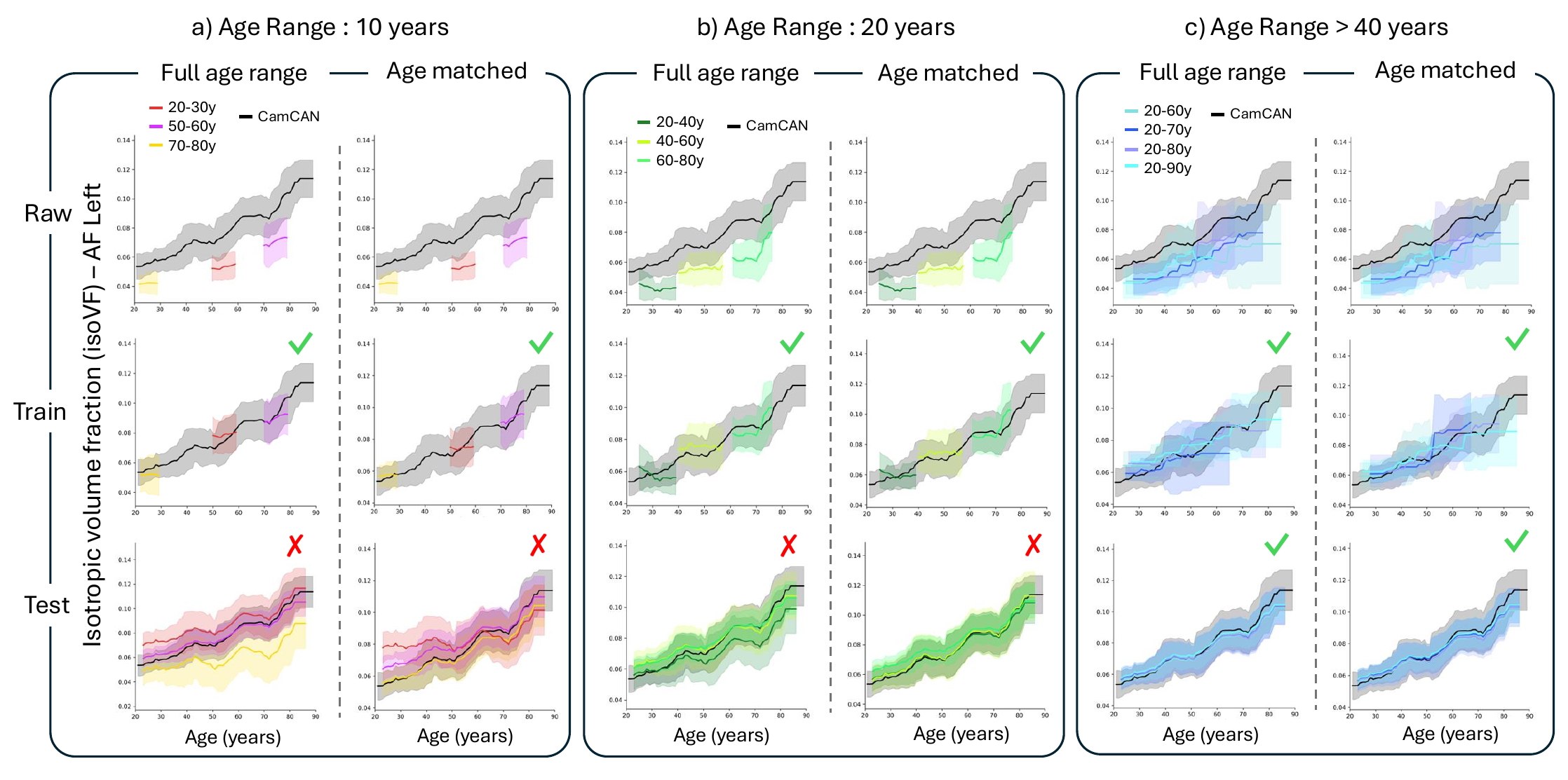} 
    \caption{This illustration is complementary to figure 6 in the paper.}
    \label{fig:S11}
\end{figure}

\begin{figure}[h]
    \centering
    \includegraphics[width=0.99\textwidth]{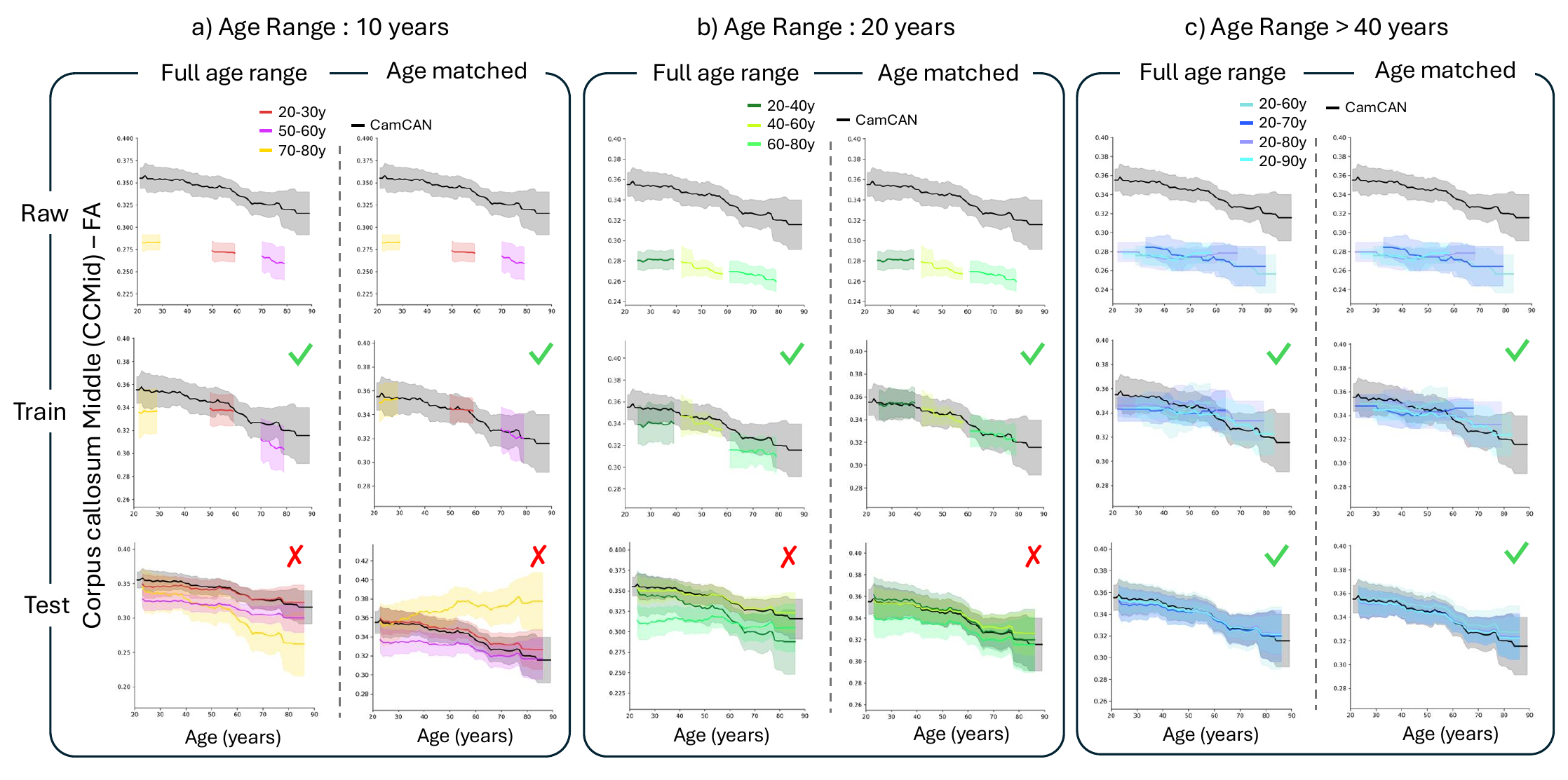} 
    \caption{This illustration is complementary to figure 6 in the paper.}
    \label{fig:S12}
\end{figure}

\begin{figure}[h]
    \centering
    \includegraphics[width=0.99\textwidth]{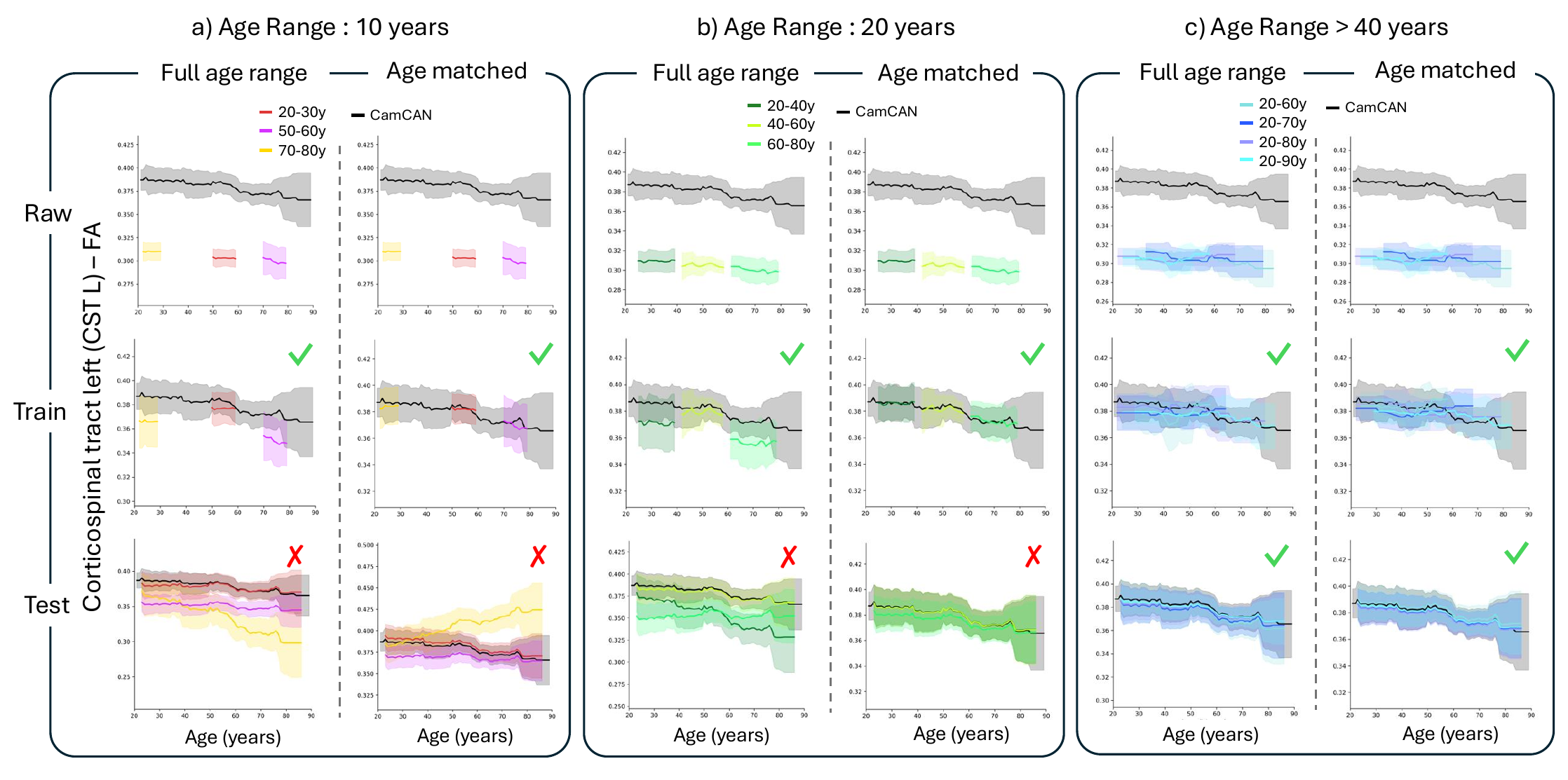} 
    \caption{This illustration is complementary to figure 6 in the paper.}
    \label{fig:S13}
\end{figure}

\begin{figure}[h]
    \centering
    \includegraphics[width=0.99\textwidth]{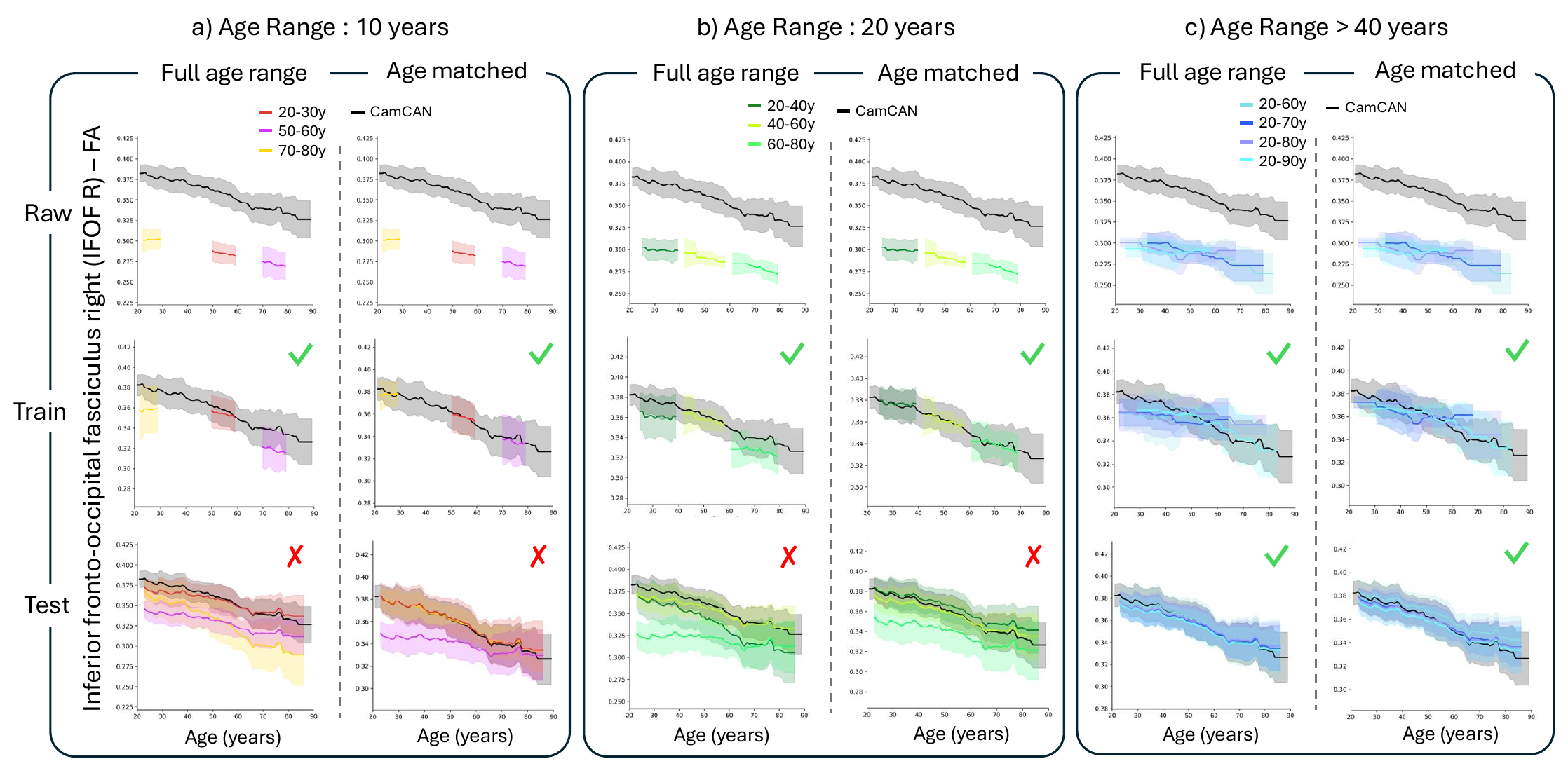} 
    \caption{This illustration is complementary to figure 6 in the paper.}
    \label{fig:S14}
\end{figure}

\paragraph{Figures \ref{fig:S9} to \ref{fig:S14} correspond to Figure 6 in the paper.} Figures \ref{fig:S9} to \ref{fig:S11} illustrate the effect of using small age ranges on the AFD, FA, and isoVF metrics measured in the AFL white matter region. Figures \ref{fig:S12} to \ref{fig:S14} show similar results for the AFD metric measured in the CC\_Mid, CST\_L, and IFOF\_R white matter regions.

\begin{figure*}[h]
    \centering
    \includegraphics[width=0.99\textwidth]{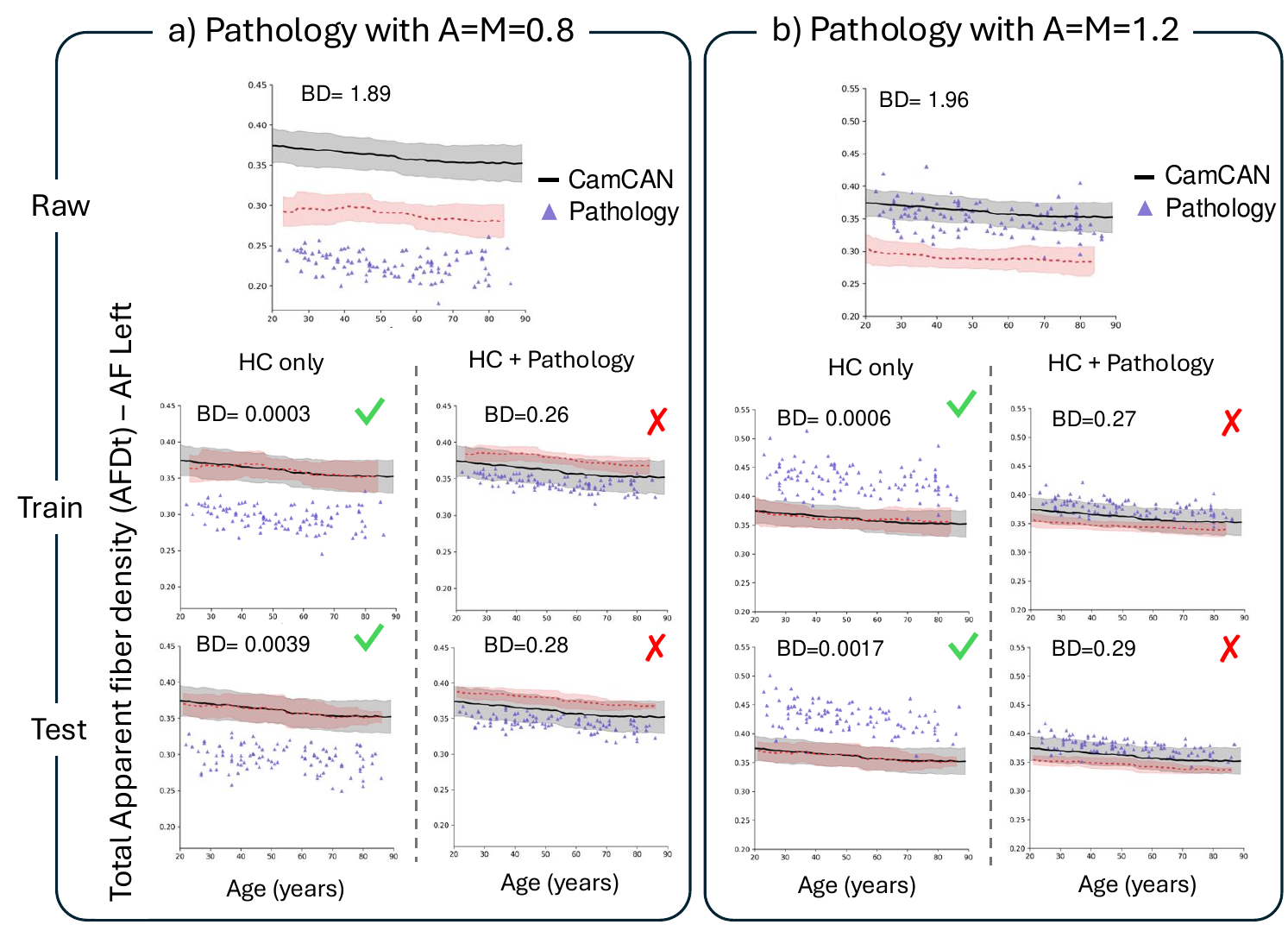} 
    \caption{This illustration is complementary to figure 9 in the paper.}
    \label{fig:S15}
\end{figure*}

\begin{figure}[h]
    \centering
    \includegraphics[width=0.99\textwidth]{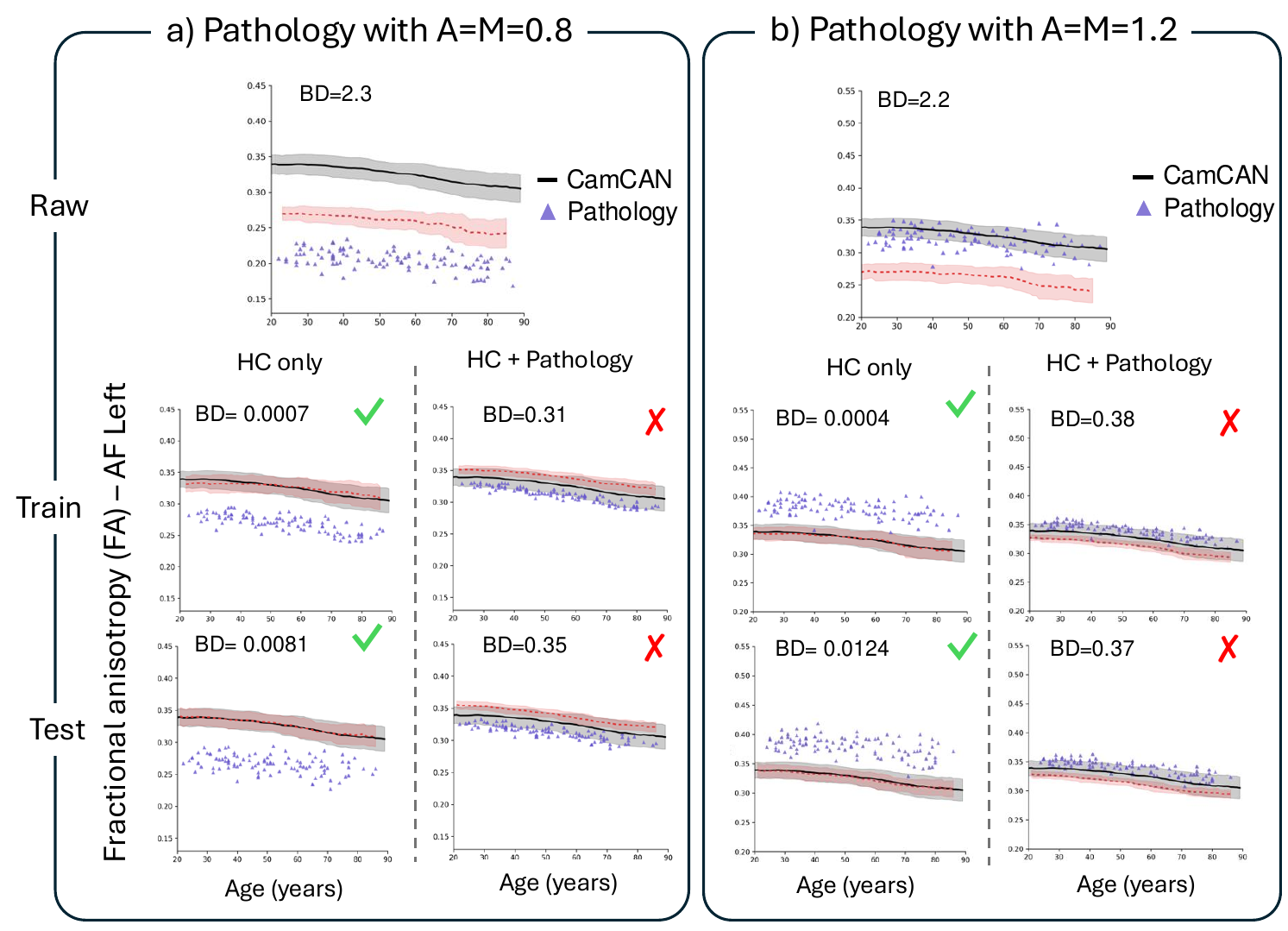} 
    \caption{This illustration is complementary to figure 9 in the paper.}
    \label{fig:S16}
\end{figure}

\begin{figure}[h]
    \centering
    \includegraphics[width=0.99\textwidth]{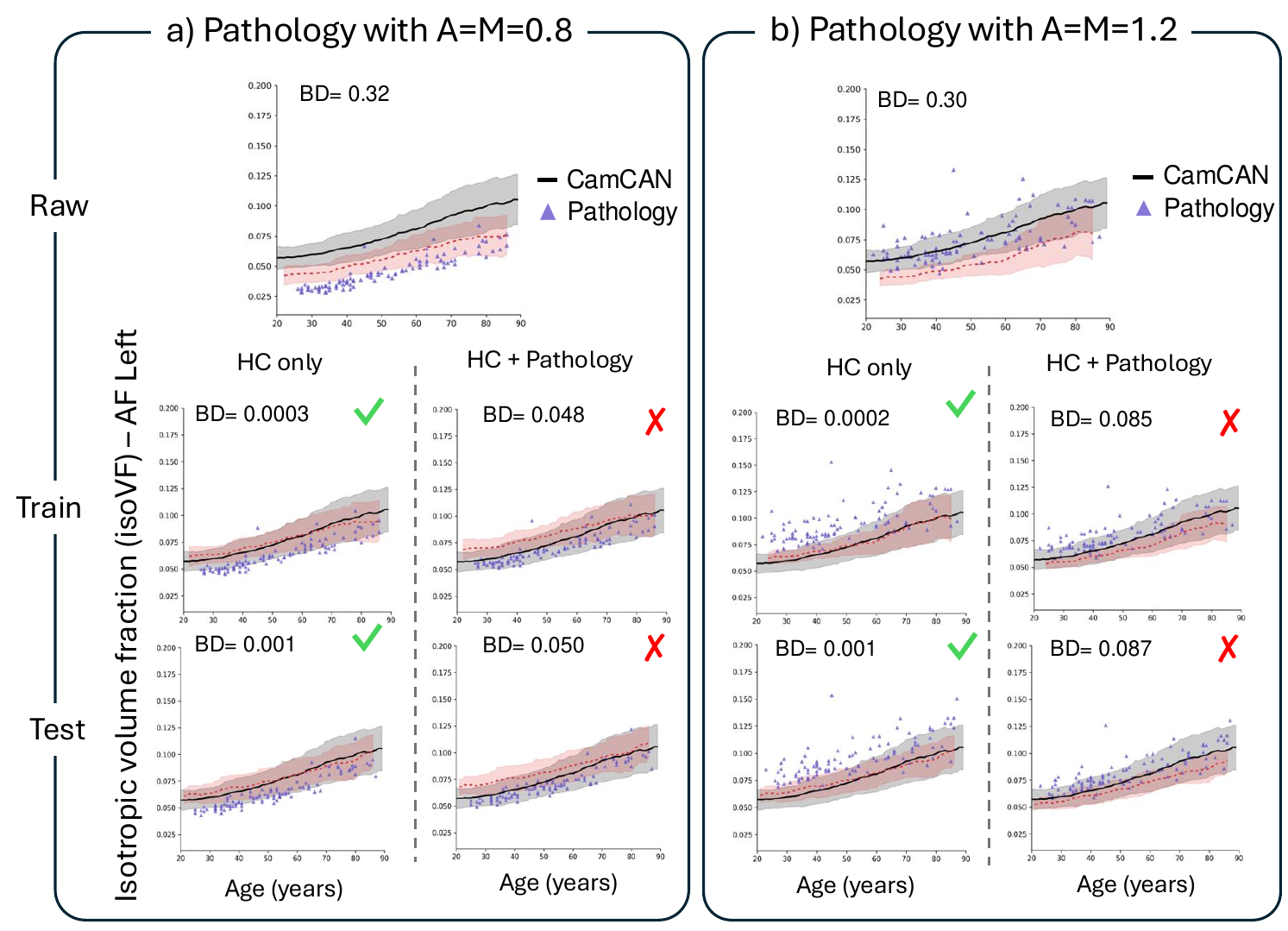} 
    \caption{This illustration is complementary to figure 9 in the paper.}
    \label{fig:S17}
\end{figure}

\begin{figure}[h]
    \centering
    \includegraphics[width=0.99\textwidth]{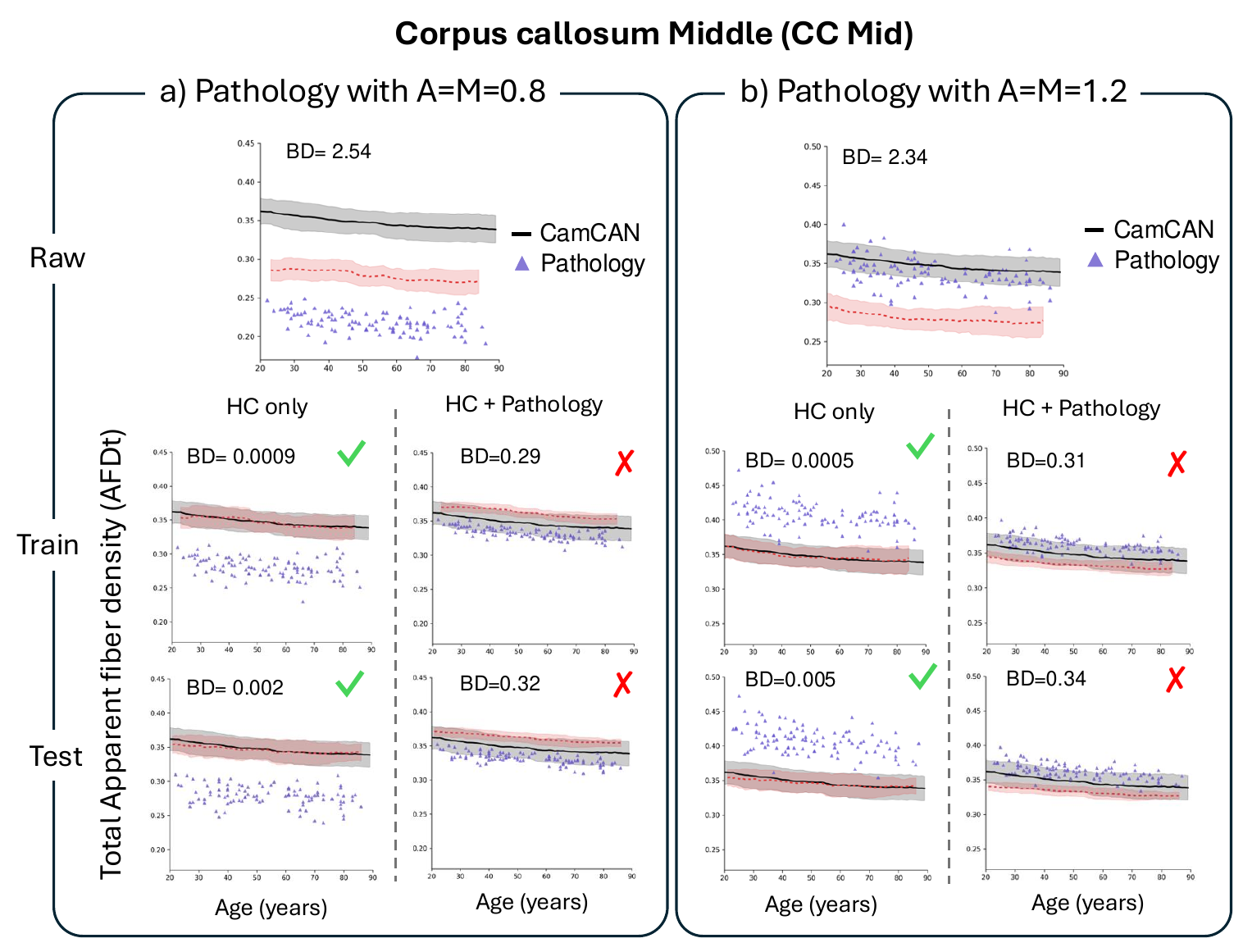} 
    \caption{This illustration is complementary to figure 9 in the paper.}
    \label{fig:S18}
\end{figure}

\begin{figure}[h]
    \centering
    \includegraphics[width=0.99\textwidth]{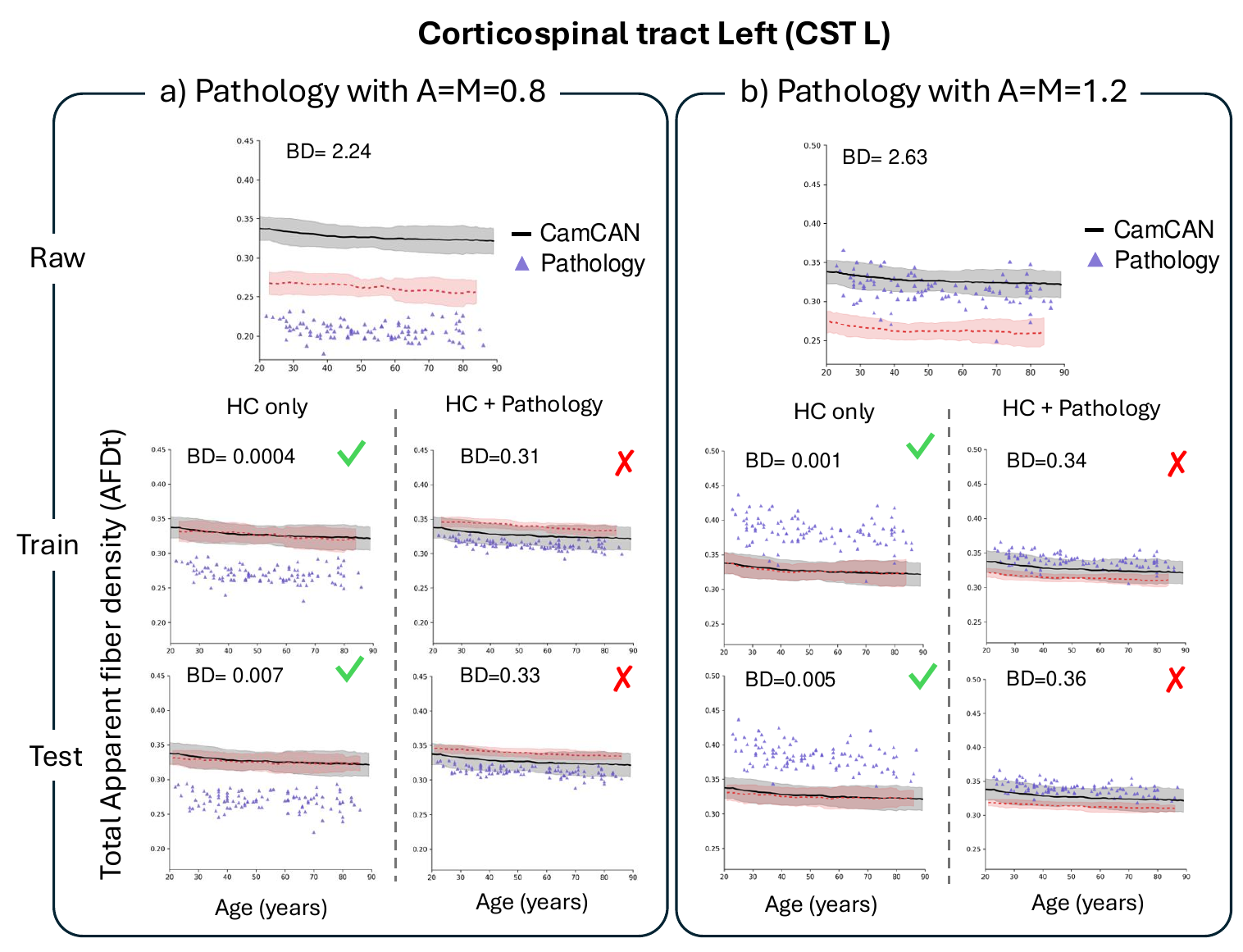} 
    \caption{This illustration is complementary to figure 9 in the paper.}
    \label{fig:S19}
\end{figure}

\begin{figure}[h]
    \centering
    \includegraphics[width=0.99\textwidth]{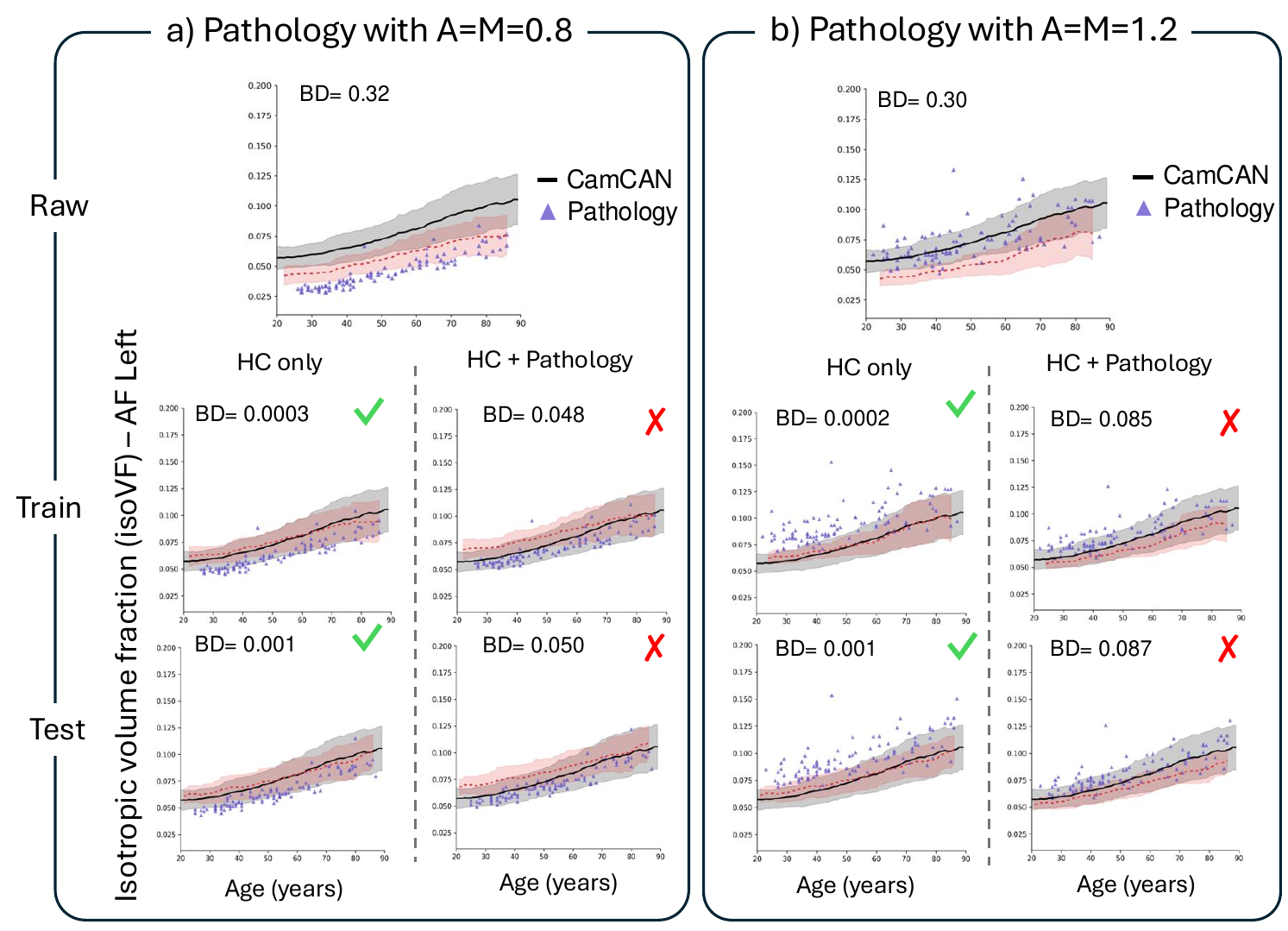} 
    \caption{This illustration is complementary to figure 9 in the paper.}
    \label{fig:S20}
\end{figure}

\paragraph{Figures \ref{fig:S15} to \ref{fig:S20} correspond to Figure 9 in the paper.} Figures \ref{fig:S15} to \ref{fig:S17} demonstrate the effect of adding pathological data to the harmonization pool when analyzing the AFD, FA, and isoVF metrics in the AFL white matter region. Figures \ref{fig:S18} to \ref{fig:S20} present similar results for the AFD metric measured in the CC\_Mid, CST\_L, and IFOF\_R white matter regions.

%%%%%%%%%%%%%%%%%%%%%%%%%%
%%%   figure 3
%%%%%%%%%%%%%%%%%%%%%%%%%%

%%%%%%%%%%%%%%%%%%%%%%%%%%
%%%   figure 5
%%%%%%%%%%%%%%%%%%%%%%%%%%

%%%%%%%%%%%%%%%%%%%%%%%%%%
%%%   figure 6
%%%%%%%%%%%%%%%%%%%%%%%%%%

%%%%%%%%%%%%%%%%%%%%%%%%%%
%%%   figure 9
%%%%%%%%%%%%%%%%%%%%%%%%%%

\end{document}